\documentclass{article}
\usepackage[margin=1.25in]{geometry}
\usepackage{natbib}
\usepackage{geometry}
\usepackage{fancyhdr}
\usepackage{afterpage}
\usepackage{graphicx}
\usepackage{amsmath,amssymb,amsbsy}
\usepackage{dcolumn,array}
\usepackage{tocloft}
\usepackage{float}
\usepackage[breaklinks=true]{hyperref}

\usepackage{amsmath}
\usepackage{amssymb,url,graphicx,booktabs,cite,epstopdf,appendix,textcomp,mathrsfs,epsfig,array}
\usepackage{graphicx}
\usepackage{amsfonts}
\usepackage{amsmath}
\usepackage{lineno}
\usepackage{placeins}

\usepackage{float}
\restylefloat{table}
\usepackage{verbatim}
\usepackage{multirow}
\usepackage{caption}
\usepackage{subfig}
\usepackage{subfloat}
\usepackage{lineno}
\usepackage{tikz} 
\usepackage{blindtext}
\usepackage{natbib}
\newcommand{\bm}{\begin{bmatrix}}
\newcommand{\fm}{\end{bmatrix}}

\usepackage{lscape}
\begin{document}

\title{The Effects of Land-use Change on Brucellosis Transmission in the Greater Yellowstone Ecosystem: A Three Species Mathematical-Epidemiological Model
}



\author{Dustin G. Padilla\medskip\\
\footnotesize{Simon A. Levin Mathematical, Computational Sciences Modeling Center}\\
 \footnotesize{Arizona State University, Tempe, AZ}}



\date{\today}
%
%

\maketitle

\begin{abstract}

This paper models brucellosis transmission between elk, cattle, and bison, of high conservation value, in the Greater Yellowstone Ecosystem. It aims to show how landscape changes in the GYE concomitantly impact brucellosis prevalence in the three species. The approach allows us to see how landscape changes in one location influence disease prevalence in populations elsewhere. The model uses the fact that the populations are configured in such a way that elk are an intermediary for disease transmission between cattle and bison. Using landscape ecology metrics applied to the habitat overlaps between elk and cattle and between elk and bison, the landscape parameters are varied to determine how disease propagates throughout the ecosystem as land-use change occurs. Result aim to provide insights into how land management used for the control of disease spread between cattle and elk, may impact (and be impacted by) disease prevalence in bison.

\end{abstract}


\section{Introduction}

Chapter 2 presented a general theoretical framework about how the shape and size of a habitat fragment and the amount of habitat overlap on a landscape impact disease transmission and persistence was developed. Chapter 3 applied this framework to brucellosis between elk and cattle in the Greater Yellowstone Ecosystem (GYE) to demonstrate its ability to approximate disease prevalence as a result of land-use change, as well as further the understanding about the relationship between patterns of landscape configuration and disease spread. In Chapter 4, the model in the previous chapter is extended to include a third species, bison, to show how landscape changes in the GYE concomitantly impact brucellosis prevalence in bison, elk, and cattle. The approach shows how landscape changes in one location influence disease prevalence in populations elsewhere. 

In ecosystems, species interact and share pathogens, and environmental change can have cascading effects on species interactions which, in turn, enables enzootic pathogens to potentially spillover to a domesticated species and cause epizootic and even zoonotic transmission \citep{patz2000effects,greer2008habitat,daszak2000emerging}.
For example, environmental change has been linked to an increase of disease spread and prevalence of swine flue, chronic wasting disease, bird flu, and other diseases in various species \citep{farnsworth2005human,wu2017economic,jones2013zoonosis,jones2008global}. As demonstrated in the previous chapter, and empirically, land-use change is one facet of environmental change that has consequences on disease spread. Land-use change is the modification of undeveloped land for anthropogenic purposes \citep{meentemeyer2012landscape}. Land-use change has been shown to alter migration patterns, species interactions, and the relative abundance of species situated in ecosystems where the land conversion occurs; moreover, land-use change has been empirically linked to higher pathogen prevalence, and more disease reservoirs throughout a landscape\citep{bradley2007urbanization,suzan2008effect,langlois2001landscape,manzione1998venezuelan,rulli2017nexus}. Although the configuration of landscape is intertwined with how diseases transmit, it is unclear how landscape changes indirectly influence disease spread in populations that are not necessarily situated on the segment of land that is modified \citep{jousimo2014ecological,suzan2012habitat}. 

Both applied and theoretical research contribute to the understanding of disease ecology. Mathematical models that study enzootics and epizootics in ecosystems have been extensively studied in plant, amphibian, and reptilian populations \citep{holdenrieder2004tree,daszak1999emerging,gibbons2000global}. Though this research indicates that land-use change is correlated with increasing disease prevalence, many models are at the global scale and not in a well-defined ecosystem \citep{crooks2011global,perrings2010globalization,gibbons2000global,jones2008global}. Moreover, the cross-species transmission models for these organisms, approximates disease prevalence well because the variety of species infected are essentially homogeneously mixed throughout the landscape, and their mixing is generally not managed anthropogenically \citep{becker2011tropical,briggs2010enzootic}. Mathematical epidemiological models to capture cross-species disease transmission between domesticated and wild species are usually between two species, and do not consider how landscape heterogeneity impact species interactions \citep{surtees1970effects,kilpatrick2009wildlife,barasona2014spatiotemporal}. That is, the models do not consider full ecological details about species interactions. Moreover, they do not consider if other species in the ecosystem are also disease reservoirs. Metapopulation models applied in epidemiological contexts are extensions of classical compartamental-epidemiological models and have incorporated aspects of spatial heterogeneity \citep{hess1996disease,mccallum2002disease}; however, these models have not explicitly considered a nested metapopulation structure for cross-species disease spread, and instead have traditionally considered fully connected metapopulations \citep{hanski2003metapopulation,gog2002disease,rowthorn2009optimal}. More specifically, they have not considered when one species mediates disease transmission between other species, and how land-use change may influence the disease prevalence directly and indirectly. Thus, this research addresses what the direct and indirect consequences of disease spread are on a species not situated on land that is converted.

To provide insight to this phenomenon, this chapter extends the system of differential equations that models brucellosis transmission between elk and cattle (discussed in the previous chapter) by including a third species, i.e. bison, and examines how landscape changes in the GYE concomitantly impact brucellosis prevalence in bison, elk, and cattle. Bison are also a reservoir for brucellosis and frequently interact with elk. Eradicating brucellosis in the region therefore requires managing it in all three species. Currently, controlling  wild species interactions is infeasible. Since elk act as intermediary for disease transmission between bison and cattle, this approach looks at how land management between elk and cattle could be used to reduce prevalence in bison, and can feed-back to lower cases in cattle. This new three species modeling approach responds to the National Academies of Sciences, Engineering, and Medicine (NASEM) report \textit{Revisiting Brucellosis in the Greater Yellowstone Area}, and adds to the existing research to understand what the indirect consequences are on disease spread in a species not situated on land that is modified, incorporates landscape ecol- \newpage \noindent ogy metrics into epidemiological models, and is the first model to consider brucellosis transmission dynamics between bison, elk, and cattle in the GYE.


\section{Background}

Brucellosis is a highly contagious mammalian bacterial infection caused by species of the genus \textit{Brucella} \citep{corbel2006brucellosis,ragan2002animal}. It transmits through contact with infected mammals' fluids, discarded contaminated matter, and vertical transmission \citep{rhyan2009pathogenesis,corbel2006brucellosis}. It can induce acute febrile, chronic gastro-intestinal dysfunction, disease-induced abortion, and a loss of fecundity in females \citep{richey1997brucella,ragan2002animal}. It affects many mammalian species, including livestock and even humans \citep{young1995overview}. Recovery from the disease can occur naturally or through treatment, but resulting immunity is often temporary \citep{corbel2006brucellosis}. Vaccination is available for various species, but also only provides temporary immunity \citep{richey1997brucella}. Globally, it is a common zoonosis and as a consequence, has many implications for trade and human health, particularly in underdeveloped socio-economic regions \citep{young1995overview,franco2007human}.

\textit{Brucella abortus} is primarily the species which infects cattle (\textit{Bos taurus}) throughout the world \citep{richey1997brucella}. In the United States, food safety measures, through the pasteurization of milk, and control efforts in cattle production have made brucellosis uncommon in people and cattle \citep{ragan2002animal}. However, in the Greater Yellowstone Ecosystem (GYE), a 12--22 million-acre region (sizes, boundaries, and descriptions of the area vary) that spans across the borders of northwestern Wyoming, southern Idaho, and southeastern Montana, \textit{B. abortus} has become endemic to wild native ungulate species and spillover cases to cattle remain persistent \citep{national2017revisiting}. 

\textit{B. abortus} was introduced to the GYE by cattle ranching in the early 1900's, but now elk (\textit{Cervus canadensis}) and bison (\textit{Bison bison}) are reservoirs of the disease \citep{meagher1994origin,cheville1998brucellosi}. The pathogen has mostly been eliminated in cattle herds through widespread vaccination, culling, and quarantine protocols \citep{ragan2002animal}; however, due to continued contact with infectious elk or with fetal remnants from disease-induced abortions, brucellosis remains persistently detected in cattle herds in the GYE \citep{national2017revisiting}. 

As brucellosis is highly contagious within herds, causes disease-induced abortion, and usually results in a loss of fecundity, its eradication is of economic concern to livestock producers in the area \citep{peck2010bovine,schumaker2012brucellosis}. Moreover, multiple agencies, such as the National Park Service (NPS), multiple branches of United States Department of Agriculture (USDA), state agencies, and tribal entities, are invested in managing the disease in the GYE for human and livestock safety, wildlife-protection, and ecological management \citep{national2017revisiting,ragan2002animal}. The three states that encompass the GYE (Idaho, Montana, and Wyoming) have each formed brucellosis surveillance regions, which together makeup the Designated Surveillance Area (DSA) specified by the \citep{national2017revisiting} report. Within the DSA, each state monitors brucellosis prevalence with the goal of reducing the economic impact for cattle producers \citep{national2017revisiting}. The comprehensive \citep{national2017revisiting} report concluded that to fully eliminate the disease in cattle, control efforts need to be focused on bison and elk as well.

There are an estimated 4,000--6,000 wild bison in Yellowstone National Park (YNP) and Grand Teton National Park (GTNP) \citep{national2017revisiting}; moreover, it is estimated that up to 60\% of the population is infected with brucellosis \citep{national2017revisiting}. Even though the disease is not considered detrimental to the bison population, their hosting of the pathogen is potentially problematic for nearby cattle herds \citep{brennan2014multi,white2011management}. Thus, bison movements are closely monitored to prevent interactions with cattle. Bison  have been restricted to the national parks in order to minimize contact with and potential brucellosis transmission to cattle on public and private lands, as the species have similar foraging ranges and have been documented to cross-species mate \citep{treanor2007brucellosis}. Although bison continue to be confined to the national parks, ecologists recommend that more habitat should be allocated outside the park for their long-term conservation \citep{scurlock2010status,baldes2016cultural}. Nevertheless, because of the management efforts to keep the species spatially separated, bison currently have not been documented to transmit the disease to cattle \citep{jones2010timing,national2017revisiting}. It has been well established, however, that bison and elk transmit brucellosis among and between their populations, as both species are reservoirs for the disease \citep{ferrari2002bison,cross2010probable,dobson1996population}.

There are an estimated 50,000 elk, separated into 9 major herds, in the GYE \citep{middleton2013animal,nelson2012elk,national2017revisiting}. They migrate from the national parks to lowland grasslands outside of the national parks in the winter to forage for food \citep{,cotterill2020disease}. Multiple herds of elk are given supplemental food resources, such as alfalfa hay, during the winter season with the goal of deterring them from interacting with cattle \citep{brennan2017shifting,cotterill2018winter}. However, it has been shown that this practice increases the population densities of elk, and, in turn, exacerbates brucellosis prevalence in their population and contributes to the pathogen's presence in cattle \citep{cotterill2018winter,cross2010probable}. Moreover, because elk migrate from the national parks to areas that are inhabited by cattle, elk act as an intermediary for disease transmission between cattle and bison, as brucellosis seroprevalence ranges from 20--60\% in elk \citep{national2017revisiting}.

The continual interaction of cattle and elk facilitates the reoccurring brucellosis incidence in cattle herds \citep{proffitt2011elk,rayl2019modeling}. Approximately 450,000 cattle in the GYE are located on lowland grasslands, both privately owned and publicly leased grazing allotments from the United States Forest Service (USFS) and Bureau of Land Management (BLM) \citep{national2017revisiting}. Most land in the GYE is publicly owned, however, most lowland grasslands are on private lands and are primarily utilized for open-grazing calf-cow cattle ranching \citep{gosnell2006ranchland,davis2011trajectories}. Moreover, it has been determined that the majority of new brucellosis infections in cattle herds are transmitted from elk on privately owned lands \citep{rayl2019modeling}. As elk migrate to the lowland grasslands in the winter, cattle herds are persistently infected with brucellosis either through coming into contact with an infectious elk or fetal remnants of disease-induced aborted fetuses \citep{brennan2017shifting,cross2010probable}. Even though a vaccination against brucellosis for cattle exists, some ranchers choose not to vaccinate their herds \citep{national2017revisiting}. Additionally, for those who do vaccinate, there are still challenges in protecting their livestock against infection; in particular, the efficacy of the vaccine does not establish complete immunity, can only first be administered when a calf is 6 months old, and needs to be re-administered every 2--3 years \citep{national2017revisiting}. Because of some lack of vaccination, limitations of the vaccine, and cattle's consistent interaction with elk, brucellosis is continually transmitted to cattle herds \citep{cross2010probable,national2017revisiting,aune2012environmental}.

Since elk act as an intermediary for brucellosis transmission between cattle and bison, it has been determined that to eradicate cases in livestock in the GYE, management efforts need to be simultaneously focused on the disease reservoirs, bison and elk \citep{cotterill2020disease,schumaker2012brucellosis}. Unfortunately, the disease control protocols for cattle are currently impractical for such wide-ranging wildlife \citep{national2017revisiting}. In theory, spatial-temporal separation of the species is the most effective control method to suppress brucellosis transmission \citep{kilpatrick2009wildlife}. However, land-use change in the GYE has increased the proximity of cattle, elk, and bison to each other \citep{national2017revisiting}. Moreover, it has altered elk migration routes such that they interact with cattle more often, and increased the relative abundance of the species in their respective habitats \citep{rickbeil2019plasticity}. 

\subsection{Research Objective}

The larger picture is that over time, land-use change 
has altered the migration patterns of elk and species interactions \citep{hansen2002ecological,davis2011trajectories}. This has facilitated brucellosis transmission throughout cattle, elk, and bison as the landscape between elk and cattle has been modified and, thus, has induced more disease reservoirs throughout the GYE \citep{national2017revisiting}. Based on these observations, this research aims to determine how land-use change between elk and cattle indirectly influences disease prevalence in bison, as well as how land-use change influences cattle brucellosis prevalence due to altered migratory patterns of elk and their interactions with bison. Moreover, the study considers how land management can be utilized to control brucellosis prevalence in these species and the implications potential polices have for vested stakeholders.

To approach these research aims, this chapter develops a metapopulation epidemiological (a three-species $SIR$ type system of differential equations) model to discern how fluctuations in landscape parameters where elk and cattle interact impact the brucellosis prevalence in bison, which are not situated on that area. All three populations are compartmentalized into susceptible, infectious, and recovered classes. The system of differential equations is set up so there is an exchange of pathogens between elk and cattle, elk and bison, but not cattle and bison. The transmission of brucellosis between different species is a function of habitat overlap, which is determined from metrics in landscape ecology. Although there are domesticated bison in the region, this study is concerned only with wild bison now constrained to national parks. 

This new approach adds to the existing disease ecology research by providing insights to the indirect consequences of disease spread in a species not situated on land that is modified. It also contributes novelty by incorporating landscape ecology metrics into mathematical epidemiological models. This study is a modeling response to the \citet{national2017revisiting} report, and considers the significance that land-use change has on disease transmission, specifically in the GYE. This research addresses implications for management and the trade-offs with the understanding that no single control measure can eliminate brucellosis from the area while completely satisfying stakeholders' needs. To our knowledge, this is the first model to consider brucellosis transmission in the GYE between bison, elk, and cattle.


\section{Three Species Brucellosis Model}

\subsection{Model Development}


The \citet{national2017revisiting} report is a peer-reviewed, comprehensive document that provides evidence-based consensus about brucellosis in the Greater Yellowstone area. The report includes findings, conclusions, and recommendations based on information gathered by a committee of experts and their deliberations. While \citep{national2017revisiting} reports multiple conclusions and recommendations, this study was particularly based on the following:
\begin{itemize}

\item ``Conclusion 1: With elk now viewed as the primary source for new cases of brucellosis in cattle and domestic bison, the committee concludes that brucellosis control efforts in the GYA will need to sharply focus on approaches that reduce transmission from elk to cattle and domestic bison."

\item ``Recommendation 1: To address brucellosis in the GYA, federal and state agencies should prioritize efforts on preventing \textit{B. abortus} transmission by elk. Modeling should be used to characterize and quantify the risk of disease transmission and spread from and among elk, which requires an understanding of the spatial and temporal processes involved in the epidemiology of the disease and economic impacts across the GYA."

\item ``Recommendation 7: The research community should address the knowledge and data gaps that impede progress in managing or reducing risk of \textit{B. abortus} transmission to cattle and domestic bison from wildlife." 

\item ``Recommendation 7A: Top priority should be placed on research to better understand brucellosis disease ecology and epidemiology in elk and bison, as such information would be vital in informing management decisions."

\item ``Recommendation 7C: Studies and assessments should be conducted to better understand the drivers of land use change and their effects on \textit{B. abortus} transmission risk."

\end{itemize}

The model developed for this study was based on research by \citet{dobson1996population}, who modeled transmission between elk and bison in the Yellowstone National Park using a parameterized system of differential equations. Even though \citet{dobson1996population} estimated prevalence levels in elk and bison, they did not incorporate the role that cattle and the habitat overlap they share with elk contribute to brucellosis transmission. For simplicity, vertical transmission is omitted. The findings of \citet{rayl2019modeling}, who determined that the majority of the transmission occurred on private ranches, motivated the incorporation of habitat overlap and land-use change into the model for brucellosis transmission between cattle, elk, and bison. Other previous research has indicated that land-use change and it's impact on ecological process is a significant issue in the region \citep{rickbeil2019plasticity}, and habitat overlap has facilitated the transmission of brucellosis in the region \citep{hansen2009species,cross2010probable}. Thus, a model for \textit{Brucella} in elk, cattle, and bison is constructed using the $SIRS$ framework for infectious diseases.  The $SIRS$ framework assumes that disease-host populations are separated into compartments to represent different stages of the epidemic. These models have been studied in the context of numerous pathogens, and could be parametrically calibrated to other epizootic and zoonotic diseases.


\subsection{Model Description}

Let there be a completely connected area $(a_E)$, called the habitat of elk, with a perimeter ($\ell_E=\ell_1$), that exists completely inside an area $(z)$, the area of the the DSA in the GYE, and the remainder of the area is $(a_C=b)$, the habitat of cattle, so that $(z-a_E=a_C)$. Inside area ($a_E$), let there be a linear distance $(d_E=d_1)$, referred to as the depth of the contact zone for cattle and elk, that is assumed homogenous around the perimeter of ($a_E$). This perimeter forms an area $(o_E=o_1)$, called the contact zone or overlap of cattle and elk. Within $(a_E)$ lies area $(c_E=c_1)$, referred to as the core habitat of elk, such that $(o_E+c_E=a_E)$. Within $(c_E)$ lies another area ($a_B$), referred to as the habitat of bison, with a perimeter ($\ell_B=\ell_2$). This area contains ($o_B$), known as the overlap for elk and bison. Within  ($o_b$) lies are ($c_B=c_1$), defined as the core habitat of bison, such that ($o_B+c_B=a_B$). The overlap for elk and bison ($o_B=o_2$) is formed by letting there be a distance ($d_B=d_2$), called the depth of the contact zone for elk and bison, from the edge of ($a_B$) to some fixed distance that is assumed homogenous around the perimeter of ($a_B$). The remainder of ($c_B$) which is not contained in ($a_B$) is denoted ($r$) and is the area inhabited by only elk. Moreover, $(o_E+r+o_B)$ is assumed to be the elk foraging range. See Figure (4.1) for an illustration of the description above.

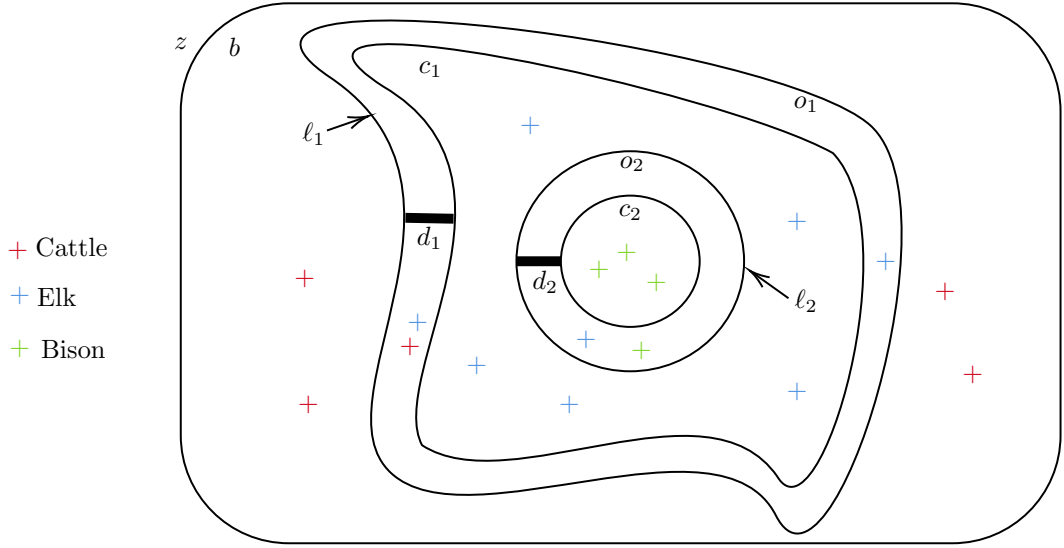
\begin{figure}[h!]
\tikzset{every picture/.style={line width=0.75pt}} 

\begin{tikzpicture}[x=0.7pt,y=0.7pt,yscale=-1,xscale=1]

\draw   (99,59) .. controls (99,26.75) and (125.15,0.6) .. (157.4,0.6) -- (516,0.6) .. controls (548.25,0.6) and (574.4,26.75) .. (574.4,59) -- (574.4,234.2) .. controls (574.4,266.45) and (548.25,292.6) .. (516,292.6) -- (157.4,292.6) .. controls (125.15,292.6) and (99,266.45) .. (99,234.2) -- cycle ;
\draw   (304.4,140.1) .. controls (304.4,120.49) and (321.19,104.6) .. (341.9,104.6) .. controls (362.61,104.6) and (379.4,120.49) .. (379.4,140.1) .. controls (379.4,159.71) and (362.61,175.6) .. (341.9,175.6) .. controls (321.19,175.6) and (304.4,159.71) .. (304.4,140.1)(280.4,140.1) .. controls (280.4,107.24) and (307.93,80.6) .. (341.9,80.6) .. controls (375.87,80.6) and (403.4,107.24) .. (403.4,140.1) .. controls (403.4,172.96) and (375.87,199.6) .. (341.9,199.6) .. controls (307.93,199.6) and (280.4,172.96) .. (280.4,140.1) ;
\draw [color={rgb, 255:red, 0; green, 0; blue, 0 }  ,draw opacity=1 ][line width=3.75]    (280.4,140.1) -- (304.4,140.1) ;
\draw   (179.6,40.6) .. controls (87.6,-21.4) and (422.6,22.6) .. (471.6,66.6) .. controls (520.6,110.6) and (450.6,336.6) .. (420.6,277.6) .. controls (390.6,218.6) and (254.6,293.6) .. (212.6,254.6) .. controls (170.6,215.6) and (271.6,102.6) .. (179.6,40.6) -- cycle ;
\draw   (208.4,45.6) .. controls (123.4,-9.4) and (390.4,47.6) .. (451.4,81.6) .. controls (493.4,125.6) and (444.4,294.6) .. (421.4,256.6) .. controls (383.4,201.6) and (277.4,270.6) .. (229.4,239.6) .. controls (209.4,201.6) and (293.4,95.6) .. (208.4,45.6) -- cycle ;
\draw [color={rgb, 255:red, 0; green, 0; blue, 0 }  ,draw opacity=1 ][line width=3.75]    (220.4,116.6) -- (246.4,117.1) ;
\draw    (178,69.4) -- (198.51,62.26) ;
\draw [shift={(200.4,61.6)}, rotate = 520.8] [color={rgb, 255:red, 0; green, 0; blue, 0 }  ][line width=0.75]    (10.93,-3.29) .. controls (6.95,-1.4) and (3.31,-0.3) .. (0,0) .. controls (3.31,0.3) and (6.95,1.4) .. (10.93,3.29)   ;
\draw    (427.4,160.1) -- (408.03,146.26) ;
\draw [shift={(406.4,145.1)}, rotate = 395.53999999999996] [color={rgb, 255:red, 0; green, 0; blue, 0 }  ][line width=0.75]    (10.93,-3.29) .. controls (6.95,-1.4) and (3.31,-0.3) .. (0,0) .. controls (3.31,0.3) and (6.95,1.4) .. (10.93,3.29)   ;

\draw (233.4,127.6) node    {$d_{1}$};
\draw (296,151) node    {$d_{2}$};
\draw (480,140.4) node  [color={rgb, 255:red, 74; green, 144; blue, 226 }  ,opacity=1 ] [align=left] {+};
\draw (527,201.4) node  [color={rgb, 255:red, 208; green, 2; blue, 27 }  ,opacity=1 ] [align=left] {+};
\draw (432,118.4) node  [color={rgb, 255:red, 74; green, 144; blue, 226 }  ,opacity=1 ] [align=left] {+};
\draw (168,217.4) node  [color={rgb, 255:red, 208; green, 2; blue, 27 }  ,opacity=1 ] [align=left] {+};
\draw (432,210.4) node  [color={rgb, 255:red, 74; green, 144; blue, 226 }  ,opacity=1 ] [align=left] {+};
\draw (166,149.4) node  [color={rgb, 255:red, 208; green, 2; blue, 27 }  ,opacity=1 ] [align=left] {+};
\draw (227,173.4) node  [color={rgb, 255:red, 74; green, 144; blue, 226 }  ,opacity=1 ] [align=left] {+};
\draw (223,186.4) node  [color={rgb, 255:red, 208; green, 2; blue, 27 }  ,opacity=1 ] [align=left] {+};
\draw (259,196.4) node  [color={rgb, 255:red, 74; green, 144; blue, 226 }  ,opacity=1 ] [align=left] {+};
\draw (309,217.4) node  [color={rgb, 255:red, 74; green, 144; blue, 226 }  ,opacity=1 ] [align=left] {+};
\draw (348,188.4) node  [color={rgb, 255:red, 126; green, 211; blue, 33 }  ,opacity=1 ] [align=left] {+};
\draw (318,182.4) node  [color={rgb, 255:red, 74; green, 144; blue, 226 }  ,opacity=1 ] [align=left] {+};
\draw (340,135.4) node  [color={rgb, 255:red, 126; green, 211; blue, 33 }  ,opacity=1 ] [align=left] {+};
\draw (356,151.4) node  [color={rgb, 255:red, 126; green, 211; blue, 33 }  ,opacity=1 ] [align=left] {+};
\draw (325,144.4) node  [color={rgb, 255:red, 126; green, 211; blue, 33 }  ,opacity=1 ] [align=left] {+};
\draw (288,66.4) node  [color={rgb, 255:red, 74; green, 144; blue, 226 }  ,opacity=1 ] [align=left] {+};
\draw (512,156.4) node  [color={rgb, 255:red, 208; green, 2; blue, 27 }  ,opacity=1 ] [align=left] {+};
\draw (128,24.4) node    {$b$};
\draw (99,22.4) node    {$z$};
\draw (234,36.4) node    {$c_{1}$};
\draw (342,113.4) node    {$c_{2}$};
\draw (437,55.4) node    {$o_{1}$};
\draw (343,87.4) node    {$o_{2}$};
\draw (171,71.4) node    {$\ell _{1}$};
\draw (437,161.4) node    {$\ell _{2}$};
\draw (11,134.4) node  [color={rgb, 255:red, 208; green, 2; blue, 27 }  ,opacity=1 ] [align=left] {+};
\draw (32,159.4) node   [align=left] {Elk};
\draw (12,158.4) node  [color={rgb, 255:red, 74; green, 144; blue, 226 }  ,opacity=1 ] [align=left] {+};
\draw (40,132.4) node   [align=left] {Cattle};
\draw (41,187.4) node   [align=left] {Bison};
\draw (12,187.4) node  [color={rgb, 255:red, 126; green, 211; blue, 33 }  ,opacity=1 ] [align=left] {+};

\end{tikzpicture}

\caption[Conceptual diagram for cattle, elk and bison habitat zones]{An illustration showing the habitats and contact zones where cattle, elk, and bison.}

\label{THREE_DIAGRAM}
\end{figure}
\newpage

Variables and parameters subscripted with capital ``$C$" are reserved for the domestic cattle population, variables and parameters subscripted with capital ``$E$" are reserved for the wild elk population, and variables and parameters subscripted with capital ``$B$" are reserved for the wild bison population. Now, suppose that there is a cattle population, $(N_C)$, homogeneously distributed, that is situated in region $(a_C+o_E)$. Also, let there exist a homogeneously distributed elk population $(N_E)$ that inhabits region $(a_E)$, and a homogeneously distributed bison population ($N_B$) that inhabits region $(a_B)$.


The value of $(z)$ represents the total acreage of the DSA, indicated by the gray region on Figure (\ref{COMPOSITE}). $(a_C+o_E)$ represents the range of all cattle herds in the DSA over the course of a year. ($a_C$) is the estimated habitat area of cattle in the DSA, which has no geographic overlap with the elk foraging range, indicated by the grey region in Figure (\ref{COMPOSITE}) that is external to the black outline. It is important to note that since the modeling approach assumes that the cattle population is homogeneously distributed throughout the DSA, it was not necessary to determine the populations' exact locations outside of the elk foraging range. The size of the contact zone between cattle and elk, $o_E$, is indicated by the purple regions in Figure (\ref{COMPOSITE}), and is estimated from global information system (GIS) data about cattle distribution on private lands and USFS and BLM grazing allotments in the DSA within the foraging range of elk. 

The habitat area of elk and bison $(a_E)$ in the GYE is the amount of hectares over which elk are located in the DSA over the course of a year, estimated from \citep{rickbeil2019plasticity} (see Figure (\ref{COMPOSITE})), and includes YNP and GTNP. The value of ($a_E$) was determined by `georeferencing' Figure (\ref{MIGRATION}) in ArcGIS and drawing a perimeter line around the span of elk migration routes estimated from \citep{rickbeil2019plasticity}, indicated by the black outline on Figure (\ref{COMPOSITE}). The  perimeter length of the black outline is $(\ell_E)$. The region of $(a_E)$ where elk do not interact with cattle is $(c_E)$, the core habitat of elk. 

Within $(c_E)$ there lies another region $(a_B)$ where bison are located. The area of bison range ($a_B$) is the combined area of YNP and GTNP, and is indicated by the green outline on Figure(\ref{COMPOSITE}) with the YNP shaded red for distinction. The perimeter length of the green outline is $(\ell_B)$. The area of elk range that does not intersect with cattle distribution nor the habitat of bison is ($r$). The area that does overlap, where the mixing of the elk and bison populations occurs, is denoted by $(o_B)$. The area of bison range that does not intersect with the elk population is ($c_B$), the core habitat of bison. Even though elk could potentially inhabit all of ($c_B$), for the purposes of modeling, the contact zone between elk and bison is assumed to be less than ($c_B$) since this abstracts data from annual elk migration.

%

%
%


\begin{figure}[h!]
\centering
{\includegraphics[scale=.6]{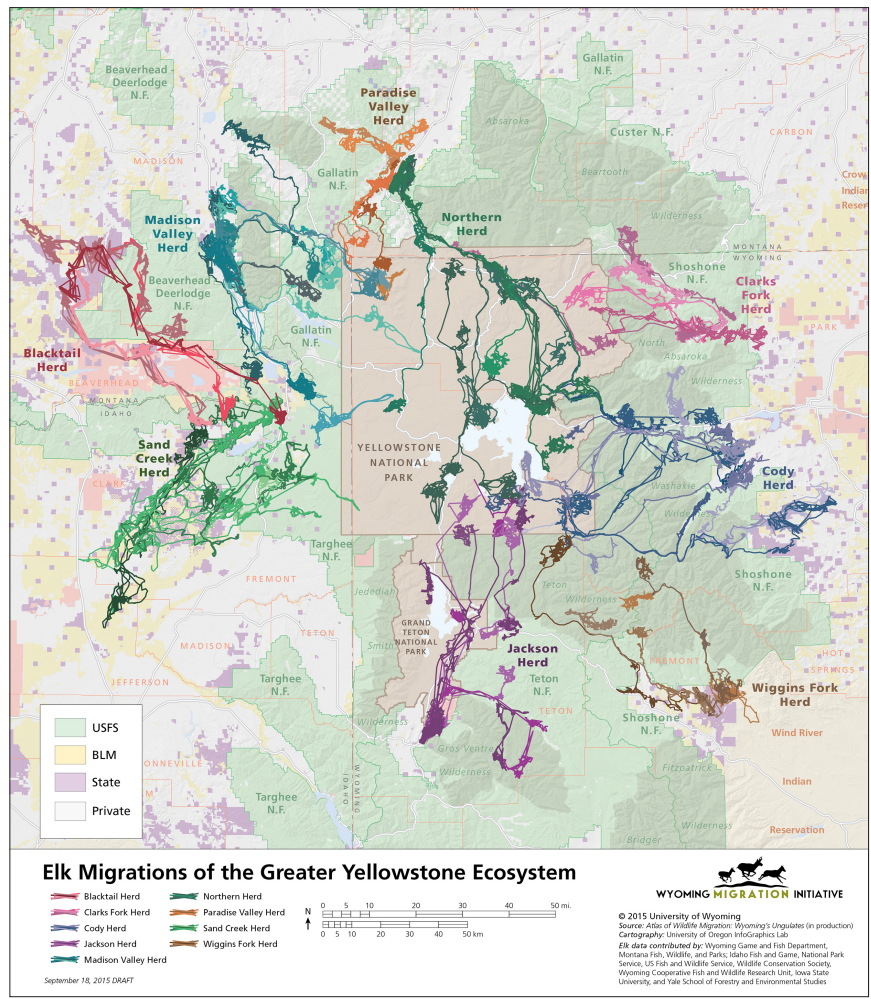}\label{MIGRATION}}
\caption[Elk Migration Map in the GYE]{Map showing various elk herds migrations in the Greater Yellowstone Ecosystem. SOURCE: \citep{rickbeil2019plasticity}
}
\label{MIGRATION}
\end{figure}


%
%
%


\begin{figure}[h!]
\centering
{\includegraphics[scale=.25]{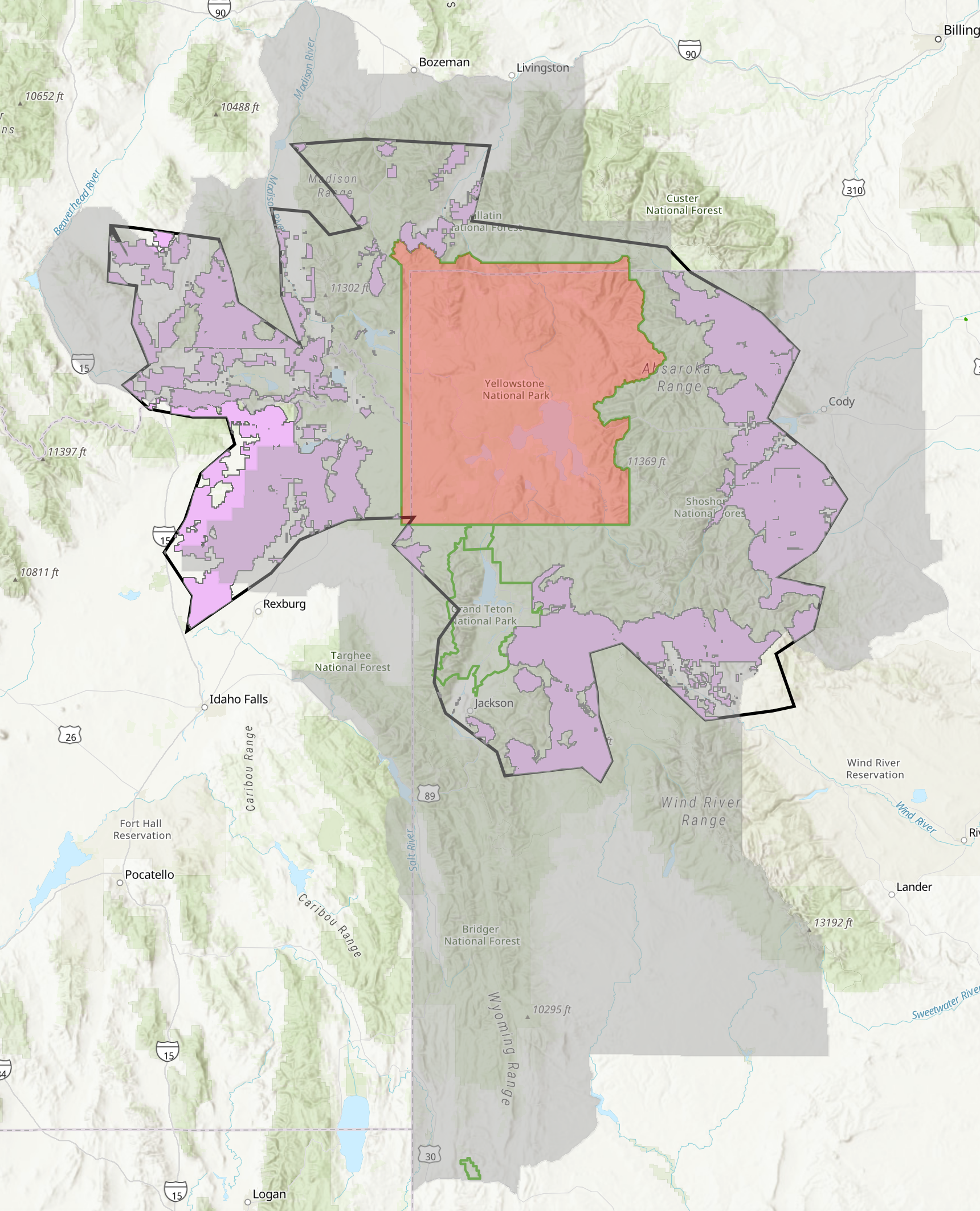}\label{COMPOSITE}}
\caption[Estimated Elk and Bison Range in GYE]{Map showing the estimated elk range (the area within the black line), the designated surveillance area (the light gray area), cattle and elk habitat overlap area (the purple areas), Yellowstone National Park (the red area), and Grand Teton National Park (the area within the green line).
}
\label{COMPOSITE}
\end{figure}


\newpage

As this model captures the spread of brucellosis amongst cattle and elk, let the cattle population $(N_C)$ be compartmentalized into classes $(S_C)$, for susceptible cattle, $(I_C)$, for infectious cattle, and $(R_C)$, for recovered cattle, so ($N_C=S_C+I_C+R_C\ge 0$) and $(S_C\ge0$), ($I_C\ge0$) and ($R_C\ge0$) for all time ($t$). As well, let the elk population $(N_E)$ be divided into classes $(S_E)$, for susceptible elk, $(I_E)$, for infectious elk, and $(R_E)$, for recovered elk, so ($N_E=S_E+I_E+R_E\ge 0$) and $(S_E\ge0$), ($I_E\ge0$) and ($R_E\ge0$) for all time ($t$).
Also let the bison population $(N_B)$ be divided into classes $(S_B)$, for susceptible elk, $(I_B)$, for infectious elk, and $(R_B)$, for recovered elk, so ($N_B=S_B+I_B+R_B\ge 0$) and $(S_B\ge0$), ($I_B\ge0$) and ($R_B\ge0$) for all time ($t$). Note that the total population of the species considered in the GYE is the sum of the subpopulations, ($N_G=N_C+N_E+N_B$).

\newpage

Assume that within-species transmission of brucellosis to cattle occurs when a susceptible cow interacts with an infectious cow in the habitat of cattle $(B)$ or on the overlap of cattle and elk $(o_E)$, and there is successful transmission. Moreover, assume that within-species transmission of brucellosis to elk occurs when a susceptible elk interacts with an infectious elk in $(r)$ or on the overlap of cattle and elk $(o_E)$ or on the overlap of elk and bison $(o_B)$, and there is successful transmission. Likewise, assume that within-species transmission of brucellosis to bison occurs when a susceptible bison interacts with an infectious bison on habitat $(c_B)$ or on the overlap $(o_B)$, and there is successful transmission. Now, assume that cross-species transmission to a susceptible member of either species occurs when it encounters an infectious individual of the another species on an overlap region, and there is successful transmission. Also, suppose an initial level of brucellosis prevalence in the infectious class of each population from literature \citep{national2017revisiting}.

Let ($b_{EC}$) be the average number of cross-species contacts that an elk makes in unit time, and let ($f_{EC}$) be the probability that cross-species contact between an infectious elk and susceptible cow transmits infection. The transmission rate of the disease from elk to cattle is then the probability a cross-species contact transmits infection times the number of contacts in unit time,
\[
\delta_{EC}=b_{EC}f_{EC}N_E.
\]
A susceptible cow has ($\delta_C$) infective interactions with the elk population in unit time, of which a fraction ($I_E$) is with infectious elk. The number of new infective cattle caused by infection from an elk in unit time is
\[
\delta_{EC}S_{C}\dfrac{I_E}{N_E}.
\]
Similarly, the transmission rate of the disease from cattle to elk is the probability that the cross-species contact transmits infection times the number of contacts in unit time,
\[
\delta_{CE}=b_{CE}f_{CE}N_C.
\]
A susceptible elk has ($\delta_{CE} $) infective interactions with the cattle population in unit time, of which a fraction ($I_C$) is with an infectious cow. The number of newly infected elk caused by infection from a cow in unit time is then
\[
\delta_{CE}S_E\dfrac{I_C}{N_C}.
\]
Likewise, the transmission rate of the disease from elk to bison is the probability that the cross-species contact transmits infection times the number of contacts in unit time,
\[
\delta_{EB}=b_{EB}f_{EB}N_B.
\]
A susceptible elk has ($\delta_{EB}$) infective interactions with the cattle population in unit time, of which a fraction ($I_B$) is with an infectious bison. The number of newly infected elk caused by infection from a bison in unit time is then
\[
\delta_{EB}S_B\dfrac{I_B}{N_B}.
\]
Lastly, the transmission rate of the disease from bison to elk is the probability that the cross-species contact transmits infection times the number of contacts in unit time,
\[
\delta_{BE}=b_{BE}f_{BE}N_B.
\]
A susceptible elk has ($\delta_{BE}$) infective interactions with the cattle population in unit time, of which a fraction ($I_B$) is with an infectious bison. The number of newly infected elk caused by infection from a bison in unit time is then
\[
\delta_{BE}S_B\dfrac{I_BE}{N_B}.
\]
Within-species infections are modeled in a similar way. Let ($b_{CC}$) be the average number of contacts that cattle and elk make with their own species in unit time, and let ($f_{CC}$) and ($f_{EE}$)  be defined as the probabilities that contact of cattle and elk with their own kind transmits infection, respectively. The transmission rate is the probability that a contact transmits infection multiplied by the number of within-species contacts in unit time for cattle is then
\[
\beta_C=b_{CC}f_{CC}N_C.
\]
The within-species transmission coefficient for elk is defined in a similar way such that 
\[
\beta_E=b_{EE}f_{EE}N_E,
\]
and the within-species transmission coefficient for bison is defined by
\[
\beta_B=b_{BB}f_{BB}N_B.
\]


After a successful disease transmission, a formerly pathogen-free animal becomes infectious and is able to infect susceptible members of its own species at rate $(\beta_C)$ for cattle, $(\beta_E)$ for elk, and  $(\beta_B)$ for bison. An infectious animal is able to cross-species infect at rate $(\delta_C)$ for cattle, $(\delta_E)$ for elk, and $(\delta_B)$ for Bison. Infectious cattle are assumed to recover at rate $(\gamma_C)$ and lose temporary immunity at rate $(\eta_C)$. Infectious elk are assumed to recover at rate $(\gamma_E)$ and lose temporary immunity at rate $(\eta_w)$. Infectious bison are assumed to recover at rate $(\gamma_B)$ and lose temporary immunity at rate $(\eta_B)$. 

All species are assumed to grow logistically, with intrinsic growth rates of ($\alpha_C$), ($\alpha_E$), and ($\alpha_E$), and have carrying capacities, ($\kappa_C$), ($\kappa_E$), and ($\kappa_E$). Also, assume that individuals in all compartments reproduce susceptible offspring. Natural mortality occurs in all compartments at rate $(\sigma_C)$ for the cattle population, $(\sigma_E)$ for the elk population, and  $(\sigma_B)$ for the bison population. Note that for this model, natural mortality of the cattle population includes ranchers' harvesting cattle for slaughter or sale. Moreover, the cattle population is assumed to grow logistically since the majority of the herd in the GYE are open-range calf-cow beef operations which would be modulated by the available rangeland. Disease induced mortality by infection of brucellosis in these populations is considered insignificant \citep{dobson1996population,national2017revisiting} and is, therefore, not included in the model. Given the description above, the brucellosis transmission model for cattle, elk and bison is then

\small
 \begin{eqnarray}\label{Model_Three}
 & & S_C' = \alpha_C N_C\left( 1-\dfrac{N_C}{\kappa_C}\right)  - \delta_C\left(\dfrac{o_E}{a_C+o_E}\right)S_C\left(\dfrac{o_E}{r+o_E+o_B}\right)\dfrac{I_E}{N_E}\nonumber\\ 
& & \text{ } \text{ }\text{ }\text{ }\text{ } \text{ }\text{ }  - \beta_C S_C\left(\dfrac{I_C}{N_C}\right) - \sigma_C S_C + \eta_C R_C, \nonumber\\
& & I_C' = \delta_C\left(\dfrac{o_E}{a_C+o_E}\right)S_C\left(\dfrac{o_E}{r+o_E+o_B}\right)\dfrac{I_E}{N_E} - \beta_C S_C\left(\dfrac{I_C}{N_C}\right) - (\sigma_C + \gamma_C) I_C, \nonumber\\
& & R_C' = \gamma_C I_C - (\eta_C + \sigma_C) R_C, \nonumber\\
 & & S_E' = \alpha_E N_E\left( 1-\dfrac{N_E}{\kappa_E}\right) - \delta_E\left(\dfrac{o_E}{r+o_E+o_B}\right)S_E\left(\dfrac{o_E}{a_C+o_E}\right)\dfrac{I_C}{N_C}\nonumber\\
& & \text{ } \text{ }\text{ }\text{ }\text{ } \text{ }\text{ }  - \delta_E\left(\dfrac{o_B}{r+o_E+o_B}\right)S_E\left(\dfrac{o_B}{c_B+o_B}\right)\dfrac{I_B}{N_B} - \beta_E S_E\left(\dfrac{I_E}{N_E}\right)+\eta_E S_E - \sigma_E R_E, \nonumber\\
& & I_E' = \delta_E\left(\dfrac{o_E}{r+o_E+o_B}\right)S_E\left(\dfrac{o_E}{a_C+o_E}\right)\dfrac{I_C}{N_C} + \delta_E\left(\dfrac{o_B}{r+o_E+o_B}\right)S_E\left(\dfrac{o_B}{c_B+o_B}\right)\dfrac{I_B}{N_B} \nonumber\\
& & \text{ } \text{ }\text{ }\text{ }\text{ } \text{ }\text{ } + \beta_E S_E\left(\dfrac{I_E}{N_E}\right) - (\sigma_E  + \gamma_E) I_E, \nonumber\\
& & R_E' = \gamma_E I_E - (\eta_E + \sigma_E) R_E, \nonumber\\
& & S_B' =  \alpha_B N_B\left( 1-\dfrac{N_B}{\kappa_B}\right) - \delta_B\left(\dfrac{o_B}{c_B+o_B}\right)S_B\left(\dfrac{o_B}{r+o_E+o_B}\right)\dfrac{I_E}{N_E}\nonumber\\
& & \text{ } \text{ }\text{ }\text{ }\text{ } \text{ }\text{ }  - \beta_B S_B\left(\dfrac{I_B}{N_B}\right) - \sigma_B S_B + \eta_B R_B, \nonumber\\
& & I_B' = \delta_B\left(\dfrac{o_B}{c_B+o_B}\right)S_B\left(\dfrac{o_B}{r+o_E+o_B}\right)\dfrac{I_E}{N_E} + \beta_B S_B\left(\dfrac{I_B}{N_B}\right) - (\sigma_B + \gamma_B) I_B, \nonumber\\
& & R_B' = \gamma_B I_B - (\eta_B + \sigma_B) R_B, \nonumber\\
& & S_C(0) = \text{ }N_C(0),\text{ } I_C(0)\text{ } =\text{ }0, \text{ }R_C(0) \text{ }=\text{ } 0, \text{ }N_C \text{ }=\text{ } S_C+I_C+R_C,\nonumber\\
& & S_E(0) = \text{ }N_E(0),\text{ } I_E(0)\text{ } =\text{ }0, \text{ }R_E(0) \text{ }=\text{ } 0, \text{ }N_E \text{ }=\text{ } S_E+I_E+R_E,\nonumber\\
& & S_B(0) = \text{ }N_B(0),\text{ } I_B(0)\text{ } =\text{ }0, \text{ }R_B(0) \text{ }=\text{ } 0, \text{ }N_B \text{ }=\text{ } S_B+I_B+R_B.
\end{eqnarray}

\normalsize

The cattle, elk, and bison populations are assumed to be homogeneously distributed throughout their respective habitats. Thus, $(\frac{o_E}{b+o_E})$ is the proportion of ($S_C$) or ($I_C$) that is in ($o_E$) at any given time, and $(\frac{o_E}{r+o_E+o_B})$ is the proportion of ($S_E$ or ($I_E$) that is in ($o_E$) at any given time. Moreover, $(\frac{o_B}{r+o_E+o_B})$ is the proportion of ($S_E$) or ($I_E$) that is in ($o_B$) at any given time, and $(\frac{o_B}{o_B+c_B})$ is the proportion of ($S_B$) or ($I_B$) that is in ($o_B$) at any given time. The size of ($o_E$) and ($o_B$) can be approximated by \[o_i = kd_i\mu_i\sqrt{a_i},\]   where    \[ \mu_i = \frac{\ell_i}{2\sqrt{\pi a_i}}\] is the shape index of each contact zone, ($\ell_i$) is the length of the perimeter of each contact zone, ($k=3.55$) is an empirically calculated scaling constant, and ($\pi=3.1415$) is the ratio of a circle's circumference to its diameter for ($i\in E,B$). The following numerical analysis assumes this approximation for the size of each contact zone.



\section{Analytical and Numerical Results}

System (\ref{Model_Three}) was mathematically analyzed using the same methods that were used for the models in Chapters 2 and 3.  A key challenge, however, is that since system (\ref{Model_Three}) is a nine-dimensional system of differential equations, the resulting analytical expressions have little biological tractability.  Even if the ($N_G'=0$), with ($N_C>0$), ($N_E>0$), and ($N_B>0$), and system (\ref{Model_Three}) is reduced to a system of six differential equations with all species surviving, the resulting analytical expressions still lack any biological interpretation. Thus, analytical results about the existence or stability conditions of the endemic (non-boundary) equilibria and the basic reproductive number are not presented. Nevertheless, some general comments about how system (\ref{Model_Three}) behaves when different subpopulations are zero are given, and the associated disease-free equilibria are discussed. Afterwards, simulations are presented to gain insight as to how solutions of system (\ref{Model_Three}) behave and biological implications are drawn based on the numerical results.

\subsection{Model Dynamics}

The components the equilibria are in the order of \[(S_C,I_C,R_C,S_E,I_E,R_E,S_B,I_B,R_B).\] The following cases address the equilibria of system (\ref{Model_Three}) in which subpopulations are equal to or greater than zero.

\begin{itemize}

\item Case 1: All subpopulations are equal to zero

\begin{itemize}
\item{}
If ($N_C=0$), ($N_B=0$), and ($N_E=0$), then the only equilibria is
 \[
 Q_0=(0,0,0,0,0,0,0,0,0).
 \]
($Q_0$) always exists and is stable if ($\alpha_C\le\sigma_C$), ($\alpha_E\le\sigma_E$), and ($\alpha_B\le\sigma_B$).

This equilibrium represents the situation when species go extinct, since the death rate for each species is greater than its birth rate. In the context of cattle production, this equilibrium implies that all individuals are culled. Again, the assumption of logistic growth is made since the majority of the cattle production in the GYE are open-range grazing operations, and elk and bison are free-roaming wild species.
\end{itemize}
\item Case 2: Two subpopulations are equal to zero
\begin{itemize}
\item{}
If ($N_i=0$), ($N_j=0$), and ($N_k>0$), for ($i,j,k$ $\in$ ($C$,$B$,$E$)), then system (\ref{Model_Three}) reduces to the classical $SIRS$ model, and has all of its dynamics \citep{hethcote2000mathematics}.
\end{itemize}
\item Case 3: One subpopulation is equal to zero

\begin{itemize}
\item{}
If either ($N_C=0$) or ($N_E=0$) with ($N_B>0$), then system (\ref{Model_Three}) reduces to the form of the model discussed in Chapter 3, and has all of its dynamics \citep{hethcote2000mathematics}.
\item{}
If ($N_C=0$) with ($N_E>0$) and ($N_B>0$), then system (\ref{Model_Three}) is decoupled into two classical $SIRS$ models, and has all of their dynamics \citep{hethcote2000mathematics}.
\end{itemize}

\item Case 4: All subpopulations are positive

\begin{itemize}

\item{}
If ($N_C>0$), ($N_E>0$), and ($N_B>0$), then system (\ref{Model_Three}) has the following (boundary) equilibrium
\[
Q_{1}=\left(\dfrac{k_C(\alpha_C-\sigma_C)}{\alpha_C},0,0,\dfrac{k_E(\alpha_E-\sigma_E)}{\alpha_E},0,0,\dfrac{k_B(\alpha_B-\sigma_B)}{\alpha_B},0,0\right)
.\]
($Q_{1}$) exists if ($\alpha_C>\sigma_C$), ($\alpha_E>\sigma_E$), and ($\alpha_B>\sigma_B$). The stability condition of ($Q_{1}$) was not able to be analytically calculated but can be shown numerically. The disease-free equilibrium is when all three species inhabit the GYE.


\item{}

In addition, if ($N_C>0$), ($N_E>0$), and ($N_B>0$), any endemic (non-boundary) equilibria of system (\ref{Model_Three}) were not able to be analytically calculated but were shown numerically. 


\end{itemize}

\end{itemize}

\subsection{Simulations}

The analysis preformed on System (\ref{Model_Three}) did not provide any expression that related brucellosis prevalence in bison to the landscape parameters between elk and cattle, ($\mu_E$) or ($d_E$); hence, the biological implications that would arise from such an expression were not able to be discovered. As a result, system (\ref{Model_Three}) was numerically solved using ode15s in MATLAB2019A. The model was parameterized from various sources, as shown in Table (4.1). 
The results are displayed with time-series plots to exemplify changes to the initial rate of infection, peak prevalence, and endemic prevalence, which are qualitative aspects of the time-series plots. The slope of a curve on a time-series plot from the initial time to the highest point on the curve indicates the initial rate of epidemic spread. The highest point of a curve on a time-series plot indicates the peak prevalence. Lastly, where a curve reaches equilibrium (flattens out) on a time-series plot indicates the endemic prevalence. Table (4.1) shows the fixed values for the simulations. The simulations help to answer questions concerning how the incorporation of a third species impacts disease transmission and how controls of landscape management could be introduced.

One question, ecologically, is how landscape alterations indirectly impact a third species. More specifically, this concerns how changes of cattle and elk interactions in their habitat overlap area impact disease prevalence in the adjacent bison population. Particularly, system (\ref{Model_Three}) was parameterized and ($\mu_E$) was varied from the shape index ($\mu_E=1$) to an expected value of ($\mu_E=300$), as discussed in Chapter 3. Additionally, the parameter space of ($d_E$) was sampled, allowing the depth of the contact zone between elk and cattle to vary.

%
%
%
%

\begin{landscape}

\begin{table}[!htbp] \begin{center}\scriptsize
\begin{tabular}{lllll} \toprule
\multicolumn{2}{l}{Parameter}              & Unit                      & Estimate  & Reference     \\ \midrule
$S_i$ & Susceptible Host Population   & cattle/ elk/ bison                            & 360,000/ 30000/ 2000 & \citep{national2017revisiting} \\
$I_i$ & Infectious Host Population       & cattle/ elk/ bison                             & 90,000/ 10,000/ 2000 & \citep{national2017revisiting} \\
$R_i$ & Recovered Host Population      & cattle/ elk/ bison                            & 0/ 0/ 0 & (Assumed) \\

$N_i$ & Host Population Size       & cattle/ elk/ bison                              &   450,000/ 40,000/ 4,000 & \citep{national2017revisiting} \\
$\alpha_i$ & Host Birth Rate      & year$^{-1}$                                 & 0.5/0.3/0.2 & \citep{national2017revisiting} \\
$\kappa_i$ & Host Carrying Capacity          & cattle/ elk/ bison                          & 500,000/ 50,000/ 10,000 & \citep{xie2009disease, dobson1996population}\\
$ \delta_{ij}$ & Cross-Species Transmission  & year$^{-1}$   &      0.3/ 0.4/ 0.5/ 0.5  & \citep{xie2009disease, dobson1996population} \\
$k$ & Scaling Constant             & unitless  &      3.55     &\citep{laurance1991predicting}        \\
$d_i$ & Depth of Contact Zone         & meter & 2.95/ 10 & \citep{rickbeil2019plasticity,USFSgrazing}$^{1}$     \\
$\ell_i$ & Contact Zone Perimeter         & meter  &     1,312,654/ 701,326 & \citep{rickbeil2019plasticity,national2017revisiting}$^{1}$     \\
$a_i$ & Species Habitat Areas      & hectare        &    5,844,938/ 9,784,751/ 1,024,552 & \citep{rickbeil2019plasticity,national2017revisiting}  \\
$z$ & Designated Surveillance Area    & hectare  &      15,629,689      & \citep{national2017revisiting}     \\
$ \beta_i $ & Within-Species Transmission     &  year$^{-1}$   & 0.3/ 0.4/ 0.5 & \citep{xie2009disease, dobson1996population}  \\
$\sigma_i$ & Host Mortality Rate         & year$^{-1}$   & 0.1/ 0.1/ 0.1 & \citep{xie2009disease, dobson1996population} \\ 
$\gamma_i$ & Loss of Immunity Rate \!\!       & year$^{-1}$  & 0.1/ 0.1/ 0.1 & \citep{xie2009disease, dobson1996population} \\
$\eta_i$ & Host Recovery Rate \!\!       & year$^{-1}$  & 0.1/ 0.1/ 0.1 & \citep{xie2009disease, dobson1996population} \\ 
$\mu_i$	  & Shape Index         & unitless    &  118/ 195                           & \citep{rickbeil2019plasticity,USFSgrazing}$^{1}$       \\ \bottomrule
\end{tabular}
\vspace{-.6cm}
\caption[Table of Parameters for Cattle-Elk-Bison Brucellosis Model]{Parameter definitions, units, and values for the model with $i\in(C,E,B)$ and $j\in(C,E,B)$. $^{1}$The value was calculated by sourcing data from USFS and BLM into ArcGISPro.}
\end{center} 
\label{BrucTable3}
\end{table}

\end{landscape}

\normalsize

\newpage

\begin{figure}[!htbp]
\centering 
\subfloat[]{{\includegraphics[scale=.085]{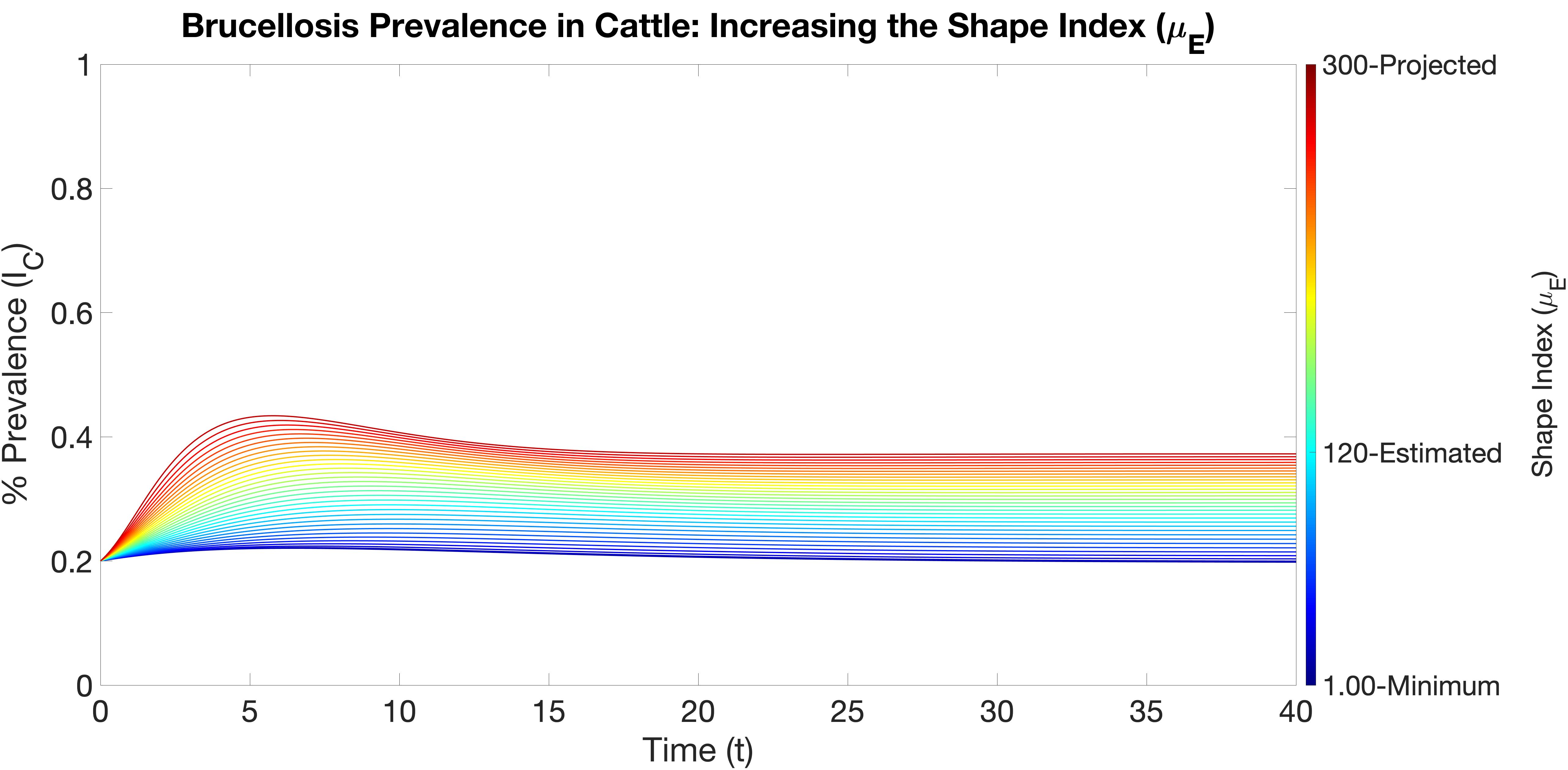}\label{THREE_CHANGE_MU1_CATTLE}}}

%

\caption
[(a) Cattle Prevalence Curves as Change Shape Index]{
(a) A time series plot showing the level of brucellosis prevalence in the cattle population as the shape index of the contact zone between cattle and elk ($\mu_E$) increases from a minimum value of ($\mu_E=1$) (the darkest blue line) to the projected value of ($\mu_E=300$) (the darkest red line). 
}
\end{figure}

\newpage

\begin{figure}[!htbp]
\ContinuedFloat
\centering 

\subfloat[]{{\includegraphics[scale=.085]{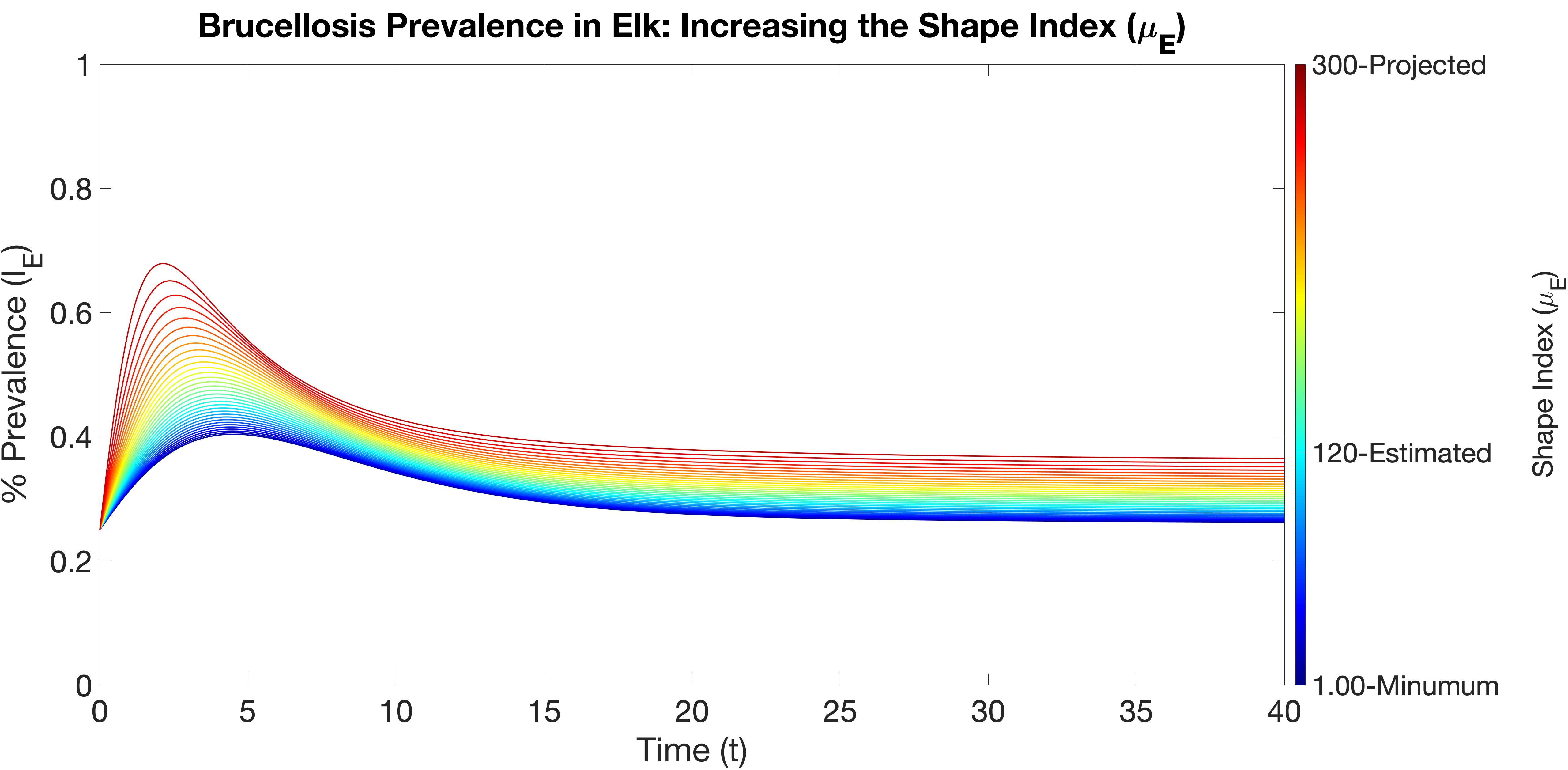}\label{THREE_CHANGE_MU1_ELK}}}


\caption
[(b) Elk Prevalence Curves as Change Shape Index]{
(b) A time series plot showing the level of brucellosis prevalence in the elk population as the shape index of the contact zone between cattle and elk ($\mu_E$) increases from a minimum value of ($\mu_E=1$) (the darkest blue line) to the projected value of ($\mu_E=300$) (the darkest red line).
}
\end{figure}

\newpage

\begin{figure}[!htbp]
\ContinuedFloat
\centering 


\subfloat[]{{\includegraphics[scale=.085]{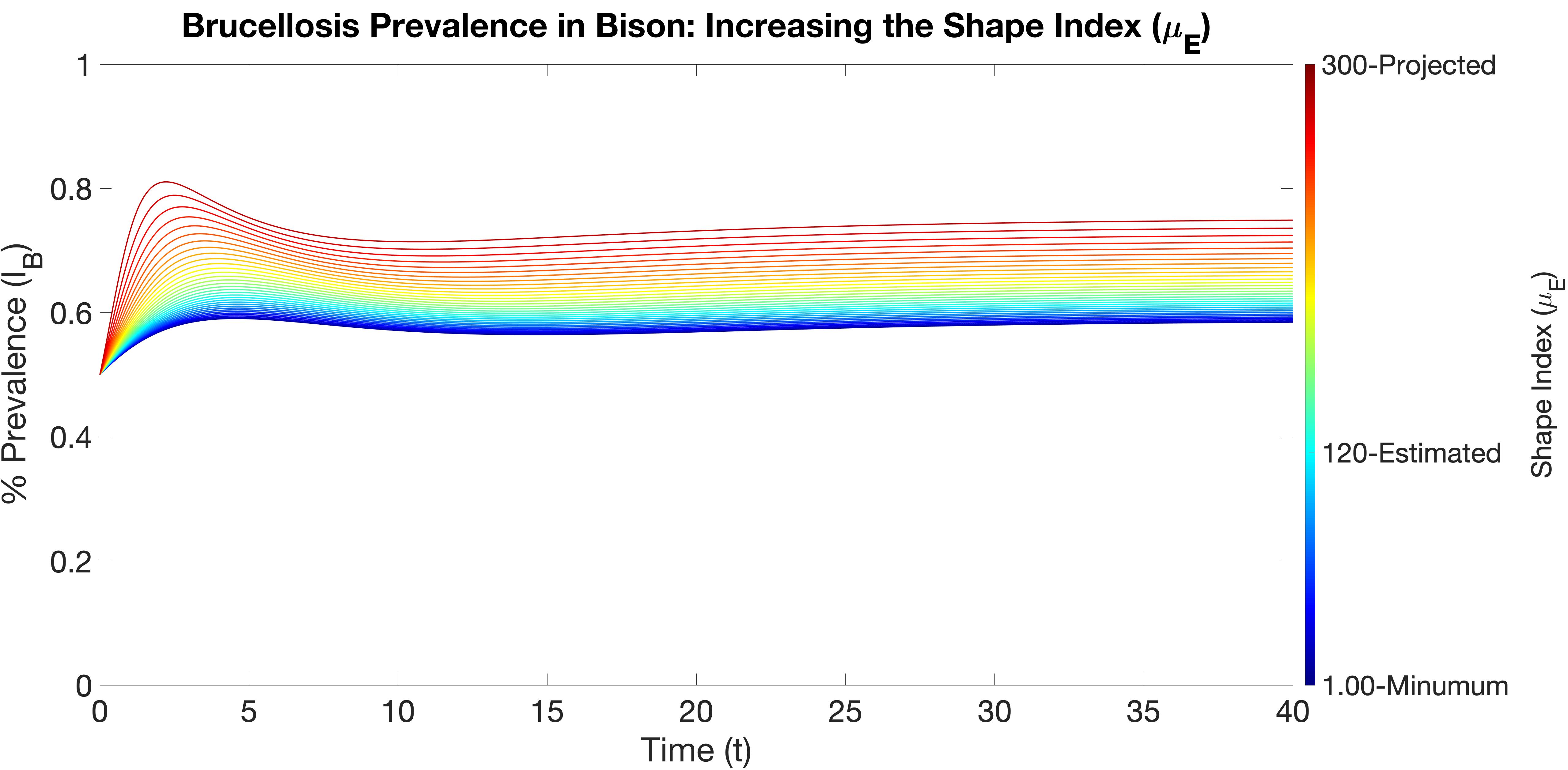}\label{THREE_CHANGE_MU1_BISON}}}

\caption
[(c) Bison Prevalence Curves as Change Shape Index]{
(c) A time series plot showing the level of brucellosis prevalence in the bison population as the shape index of the contact zone between cattle and elk ($\mu_E$) increases from a minimum value of ($\mu_E=1$) (the darkest blue line) to the projected value of ($\mu_E=300$) (the darkest red line).
}
\label{THREE_CHANGE_MU1}
\end{figure}


\newpage

\begin{figure}[!htbp]
\centering
\subfloat[]{{\includegraphics[scale=.085]{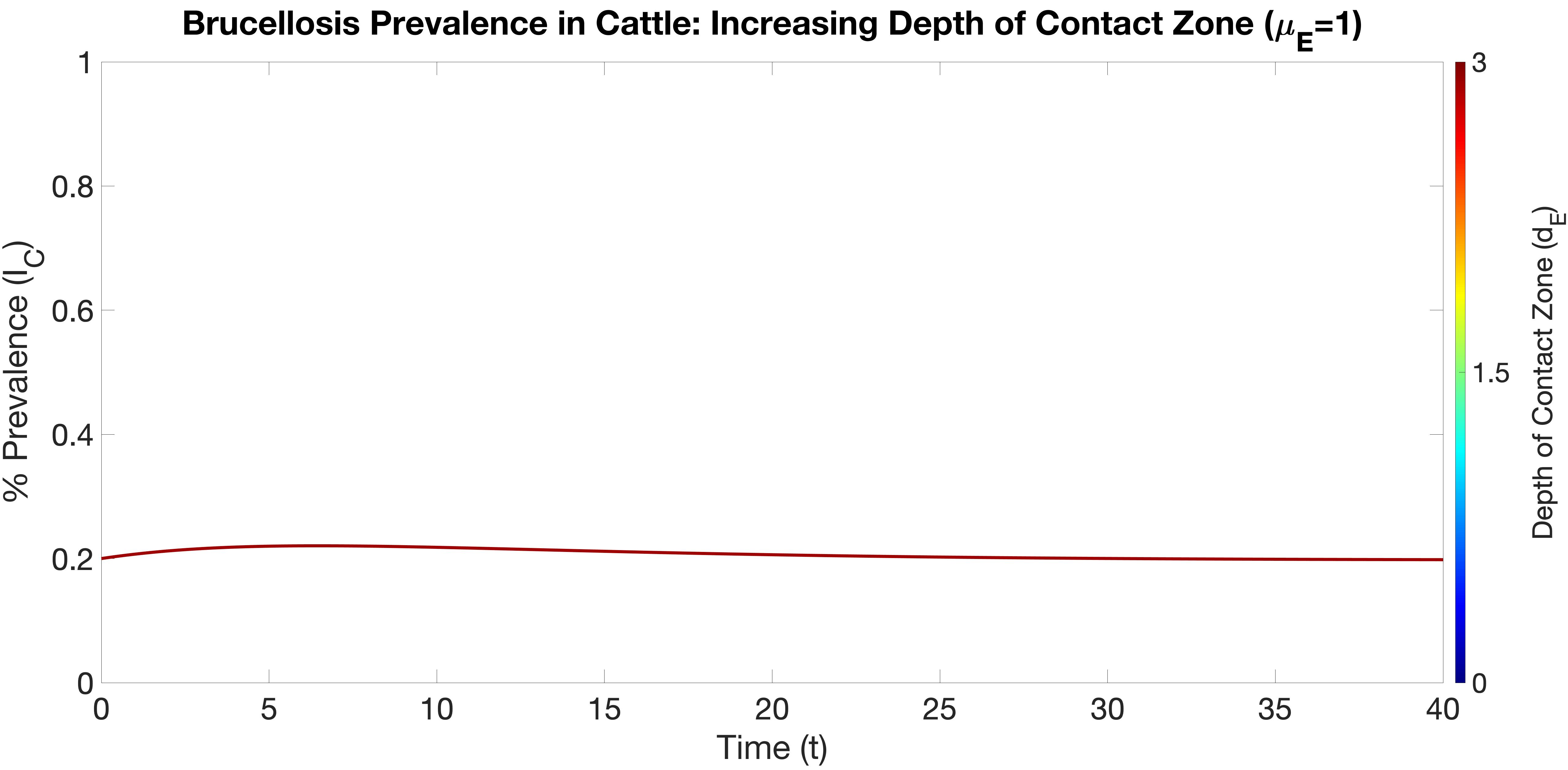}\label{THREE_CHANGE_D1_MU1_1_CATTLE}}}


\caption
[(a) Cattle Prevalence Curves as Change Depth of Contact Zone Between Cattle and Elk with Low Shape Index]{
(a) A time series plot showing the level of brucellosis prevalence in the cattle population on a landscape where the contact zone between cattle and elk has the shape index of $(\mu_E=1)$ as the depth of the contact zone ($d_E$) increases from a minimum value of ($d_E=0$) to an assigned value of ($d_E=3$) (the darkest red line which overlays the other curves).
}
\label{THREE_CHANGE_D1_MU1_1}
\end{figure}

\newpage

\begin{figure}[!htbp]
\ContinuedFloat
\centering

\subfloat[]{{\includegraphics[scale=.085]{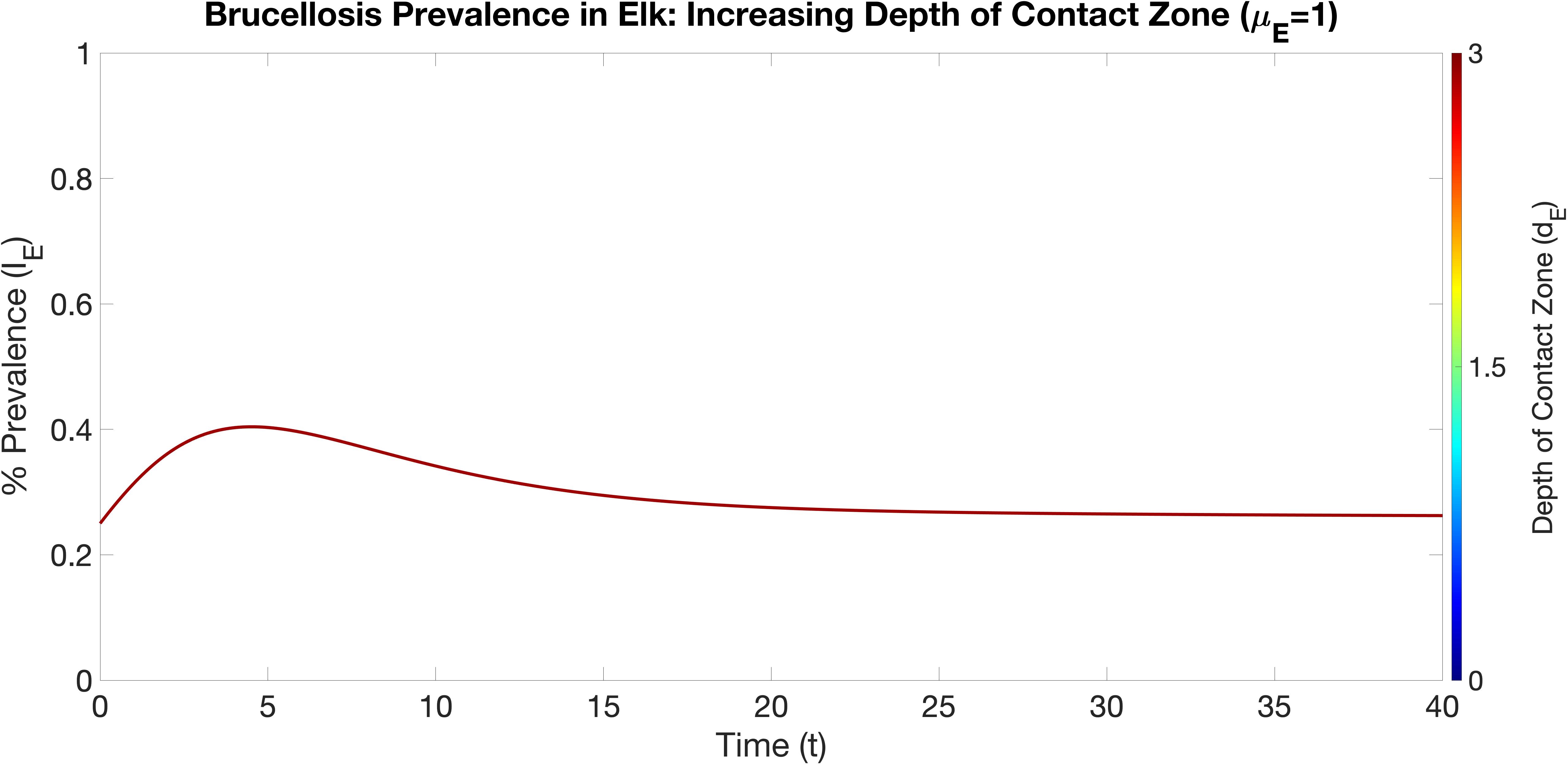}\label{THREE_CHANGE_D1_MU1_1_ELK}  }}

\caption
[(b) Elk Prevalence Curves as Change Depth of Contact Zone Between Cattle and Elk with Low Shape Index]{
(b) A time series plot showing the level of brucellosis prevalence in the elk population on a landscape where the contact zone between cattle and elk has the shape index of $(\mu_E=1)$ as the depth of the contact zone ($d_E$) increases from a minimum value of ($d_E=0$) to an assigned value of ($d_E=3$) (the darkest red line which overlays the other curves).
}
\label{THREE_CHANGE_D1_MU1_1}
\end{figure}
\newpage

\begin{figure}[!htbp]
\ContinuedFloat
\centering


\subfloat[]{{\includegraphics[scale=.085]{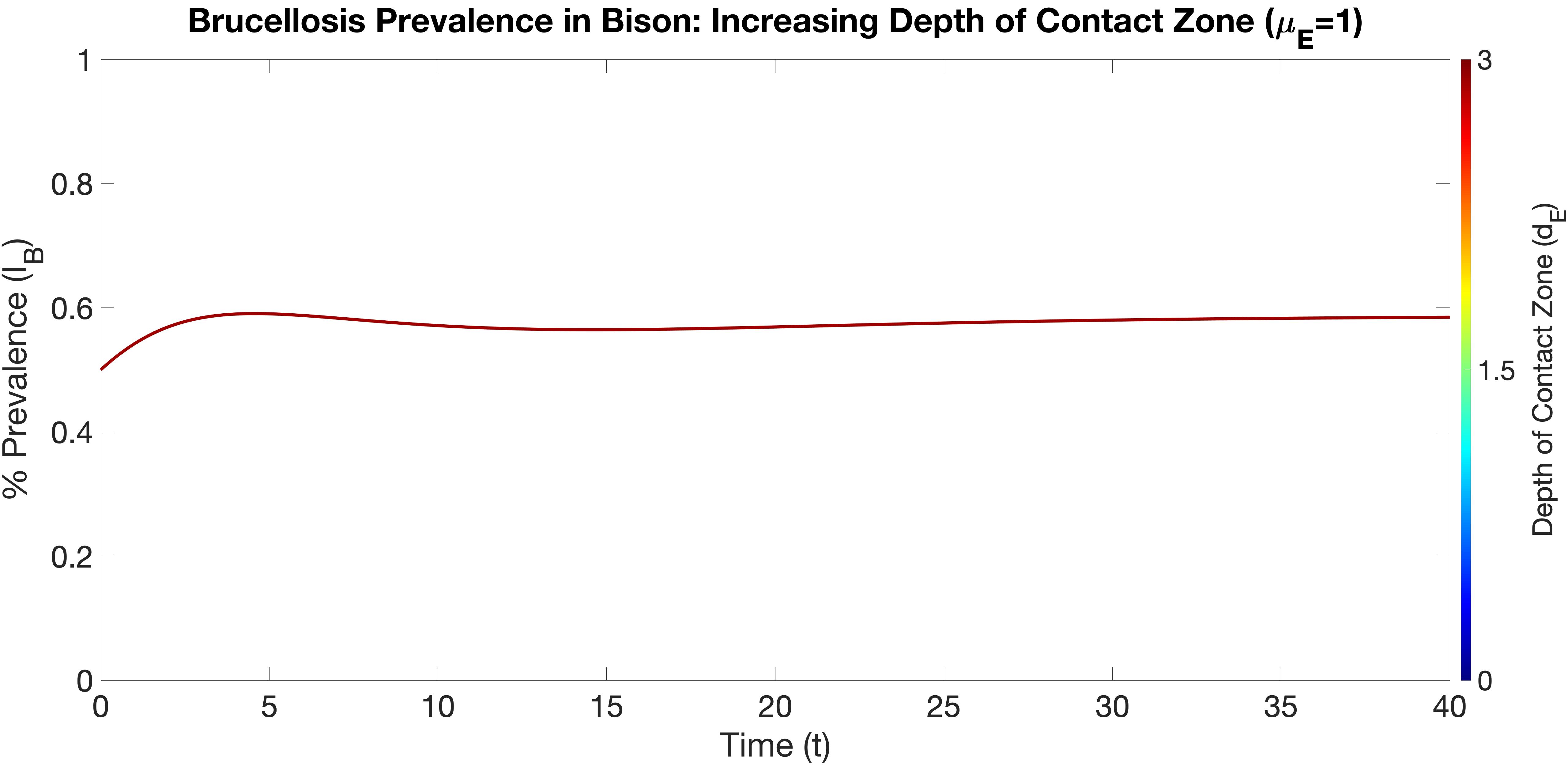}\label{THREE_CHANGE_D1_MU1_1_BISON}  }}
\caption
[(c) Bison Prevalence Curves as Change Depth of Contact Zone Between Cattle and Elk with Low Shape Index]{
(c) A time series plot showing the level of brucellosis prevalence in the bison population on a landscape where the contact zone between cattle and elk has the shape index of $(\mu_E=1)$ as the depth of the contact zone ($d_E$) increases from a minimum value of ($d_E=0$) (the darkest blue line) to an assigned value of ($d_E=3$) (the darkest red line which overlays the other curves).
}
\label{THREE_CHANGE_D1_MU1_1}
\end{figure}


\newpage

\begin{figure}[!htbp]
\centering
\subfloat[]{{\includegraphics[scale=.085]{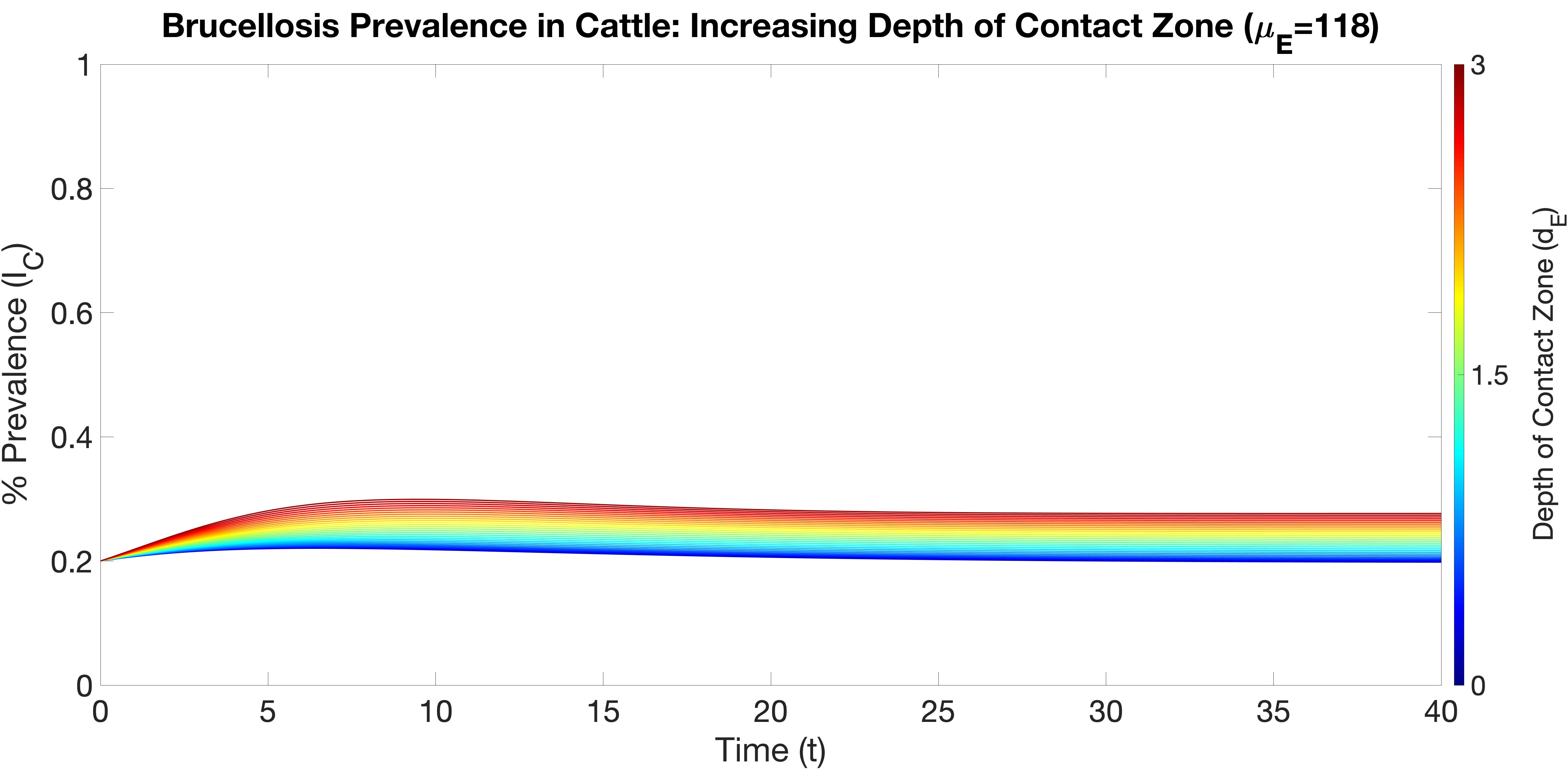}\label{THREE_CHANGE_D1_MU1_118_Cattle}   }}



\caption
[(a) Cattle Prevalence Curves as Change Depth of Contact Zone Between Cattle and Elk with Medium Shape Index]{
(a) A time series plot showing the level of brucellosis prevalence in the cattle population on a landscape where the contact zone between cattle and elk has the shape index of $(\mu_E=118)$ as the depth of the contact zone ($d_E$) increases from a minimum value of ($d_E=0$) (the darkest blue line) to an assigned value of ($d_E=3$) (the darkest red line). 
}
\label{THREE_CHANGE_D1_MU1_118}
\end{figure}
\newpage

\begin{figure}[!htbp]
\ContinuedFloat
\centering

\subfloat[]{{\includegraphics[scale=.085]{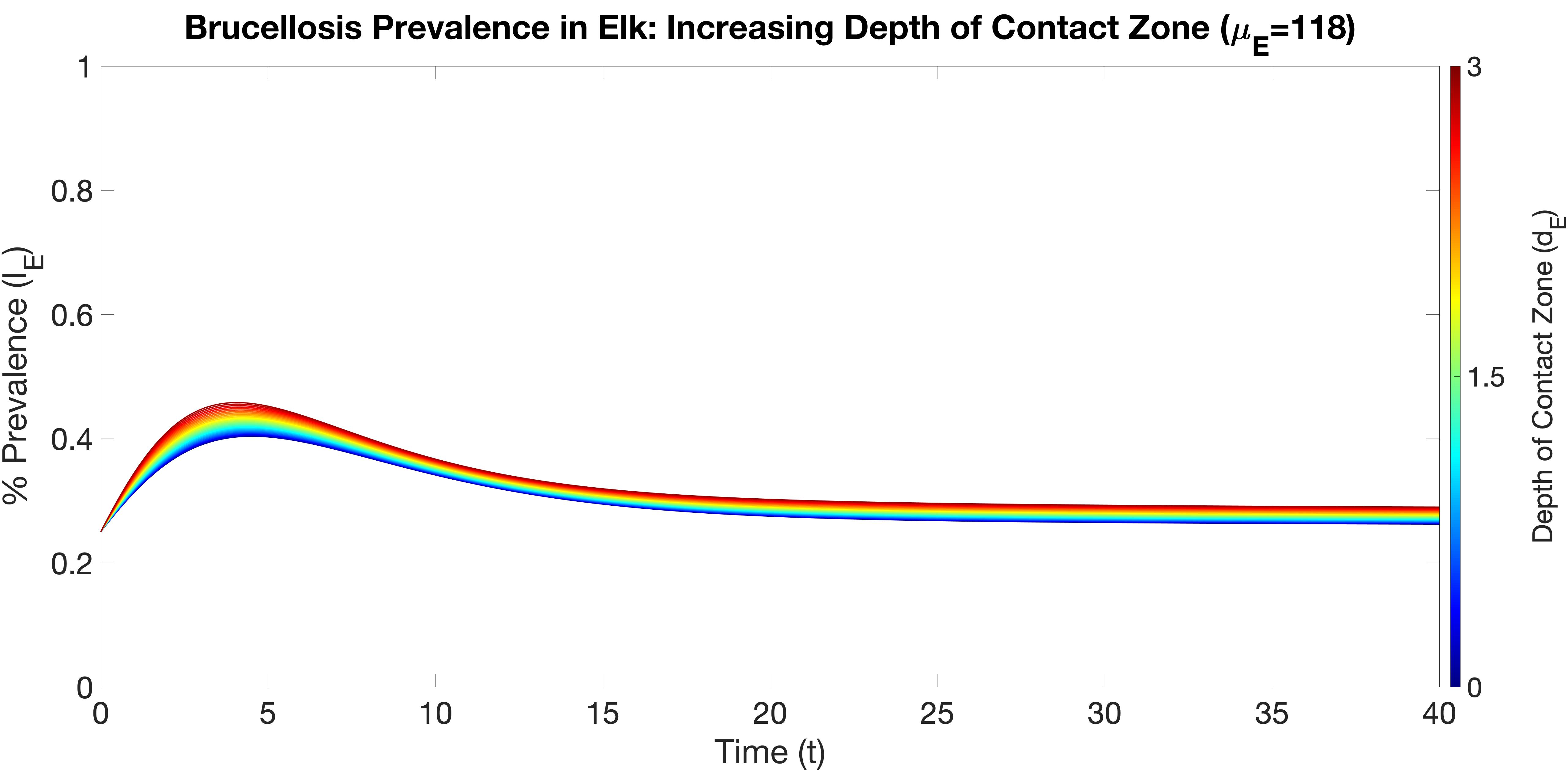}\label{THREE_CHANGE_D1_MU1_118_Elk}  }}


\caption
[(b) Elk Prevalence Curves as Change Depth of Contact Zone Between Cattle and Elk with Medium Shape Index]{
(b) A time series plot showing the level of brucellosis prevalence in the elk population on a landscape where the contact zone between cattle and elk has the shape index of $(\mu_E=118)$ as the depth of the contact zone ($d_E$) increases from a minimum value of ($d_E=0$) (the darkest blue line) to an assigned value of ($d_E=3$) (the darkest red line). 
}
\label{THREE_CHANGE_D1_MU1_118}
\end{figure}

\newpage

\begin{figure}[!htbp]
\ContinuedFloat
\centering


\subfloat[]{{\includegraphics[scale=.085]{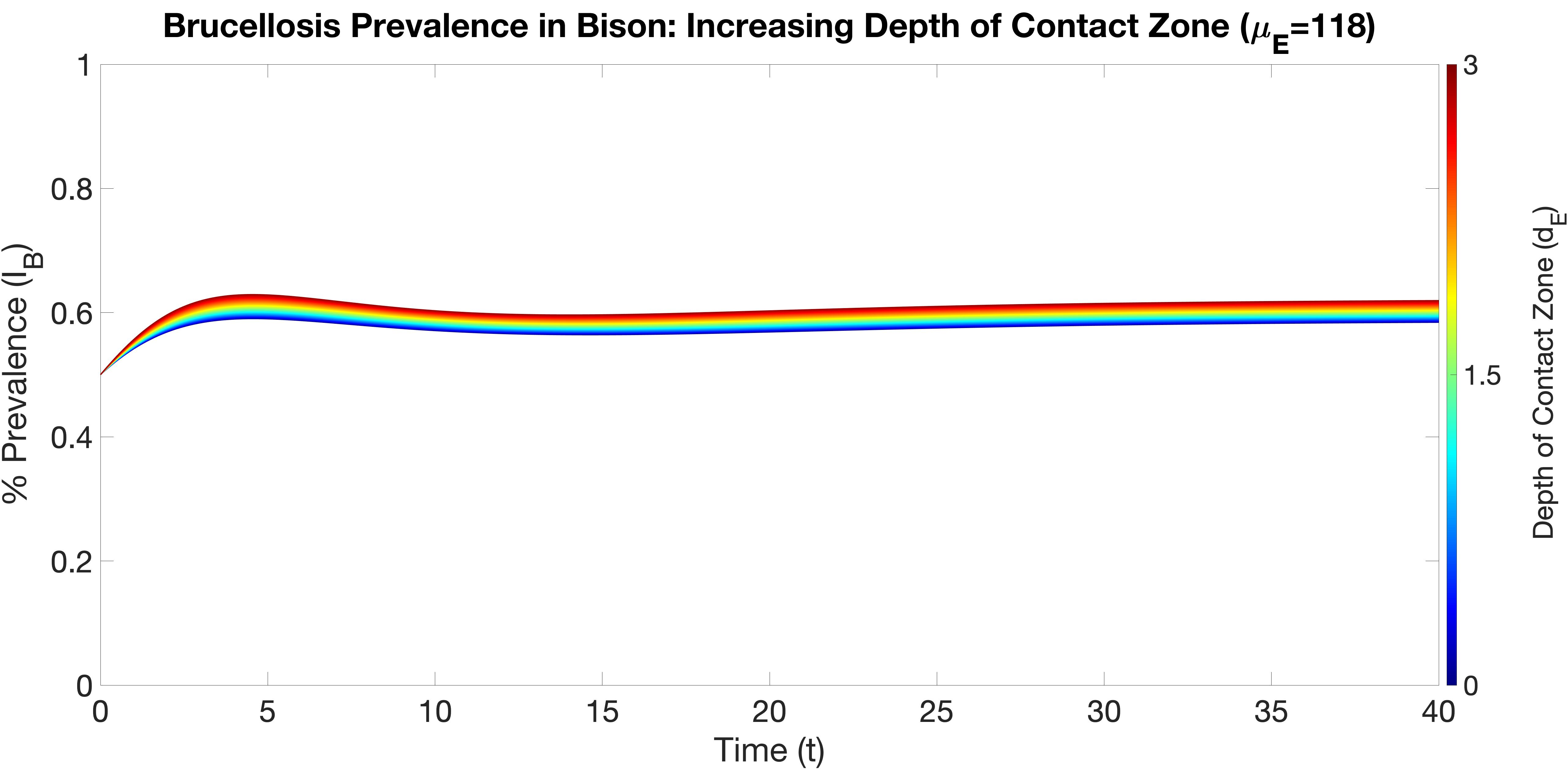}\label{THREE_CHANGE_D1_MU1_118_Bison}  }}

\caption
[(c) Bison Prevalence Curves as Change Depth of Contact Zone Between Cattle and Elk with Medium Shape Index]{
(c) A time series plot showing the level of brucellosis prevalence in the bison population on a landscape where the contact zone between cattle and elk has the shape index of $(\mu_E=118)$ as the depth of the contact zone ($d_E$) increases from a minimum value of ($d_E=0$) (the darkest blue line) to an assigned value of ($d_E=3$) (the darkest red line).
}
\label{THREE_CHANGE_D1_MU1_118}
\end{figure}


\newpage

\begin{figure}[!htbp]
\centering
\subfloat[]{{\includegraphics[scale=.085]{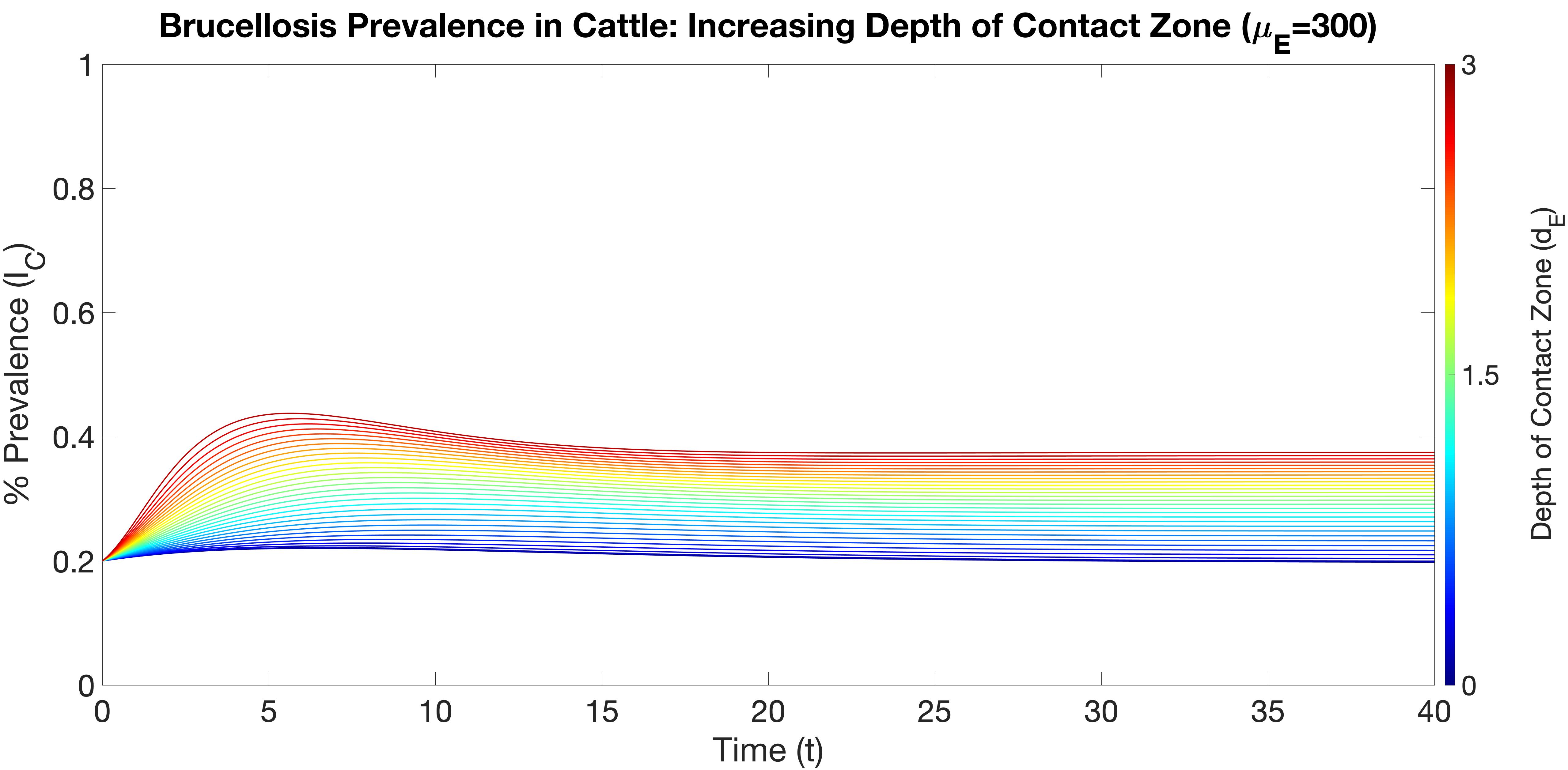}\label{THREE_CHANGE_D1_MU1_300_Cattle}   }}

%

\caption
[(a) Cattle Prevalence Curves as Change Depth of Contact Zone Between Cattle and Elk with High Shape Index]
{
(a) A time series plot showing the level of brucellosis prevalence in the cattle population on a landscape where the contact zone between cattle and elk has the shape index of $(\mu_E=300)$ as the depth of the contact zone ($d_E$) increases from a minimum value of ($d_E=0$) (the darkest blue line) to an assigned value of ($d_E=3$) (the darkest red line). 
}
\label{THREE_CHANGE_D1_MU1_300}
\end{figure}

\newpage

\begin{figure}[!htbp]
\ContinuedFloat
\centering

\subfloat[]{{\includegraphics[scale=.085]{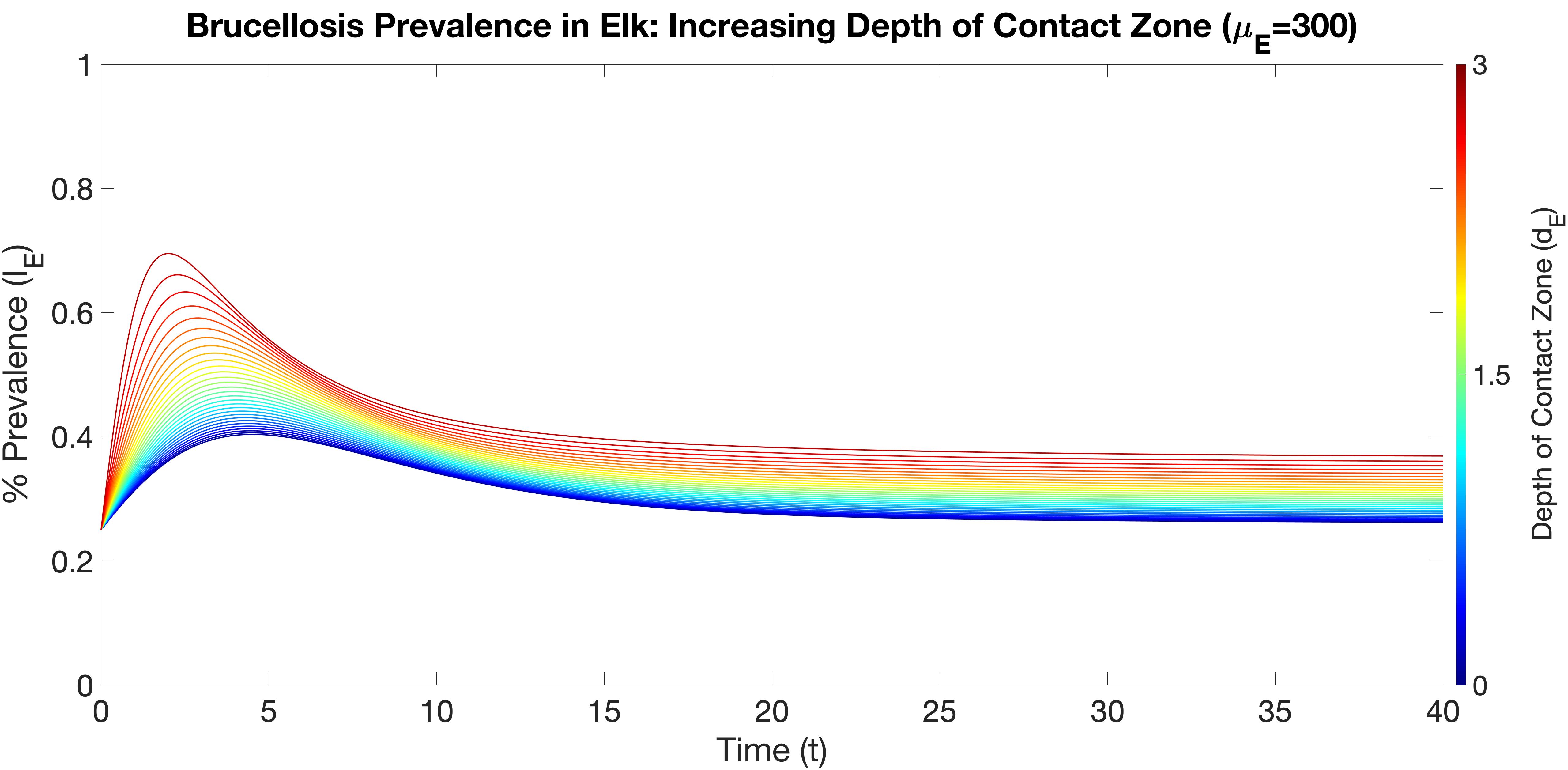}\label{THREE_CHANGE_D1_MU1_300_Elk}  }}


\caption
[(b) Elk Prevalence Curves as Change Depth of Contact Zone Between Cattle and Elk with High Shape Index]
{
(b) A time series plot showing the level of brucellosis prevalence in the elk population on a landscape where the contact zone between cattle and elk has the shape index of $(\mu_E=300)$ as the depth of the contact zone ($d_E$) increases from a minimum value of ($d_E=0$) (the darkest blue line) to an assigned value of ($d_E=3$) (the darkest red line). 
}
\label{THREE_CHANGE_D1_MU1_300}
\end{figure}

\newpage

\begin{figure}[!htbp]
\ContinuedFloat
\centering
%

\subfloat[]{{\includegraphics[scale=.085]{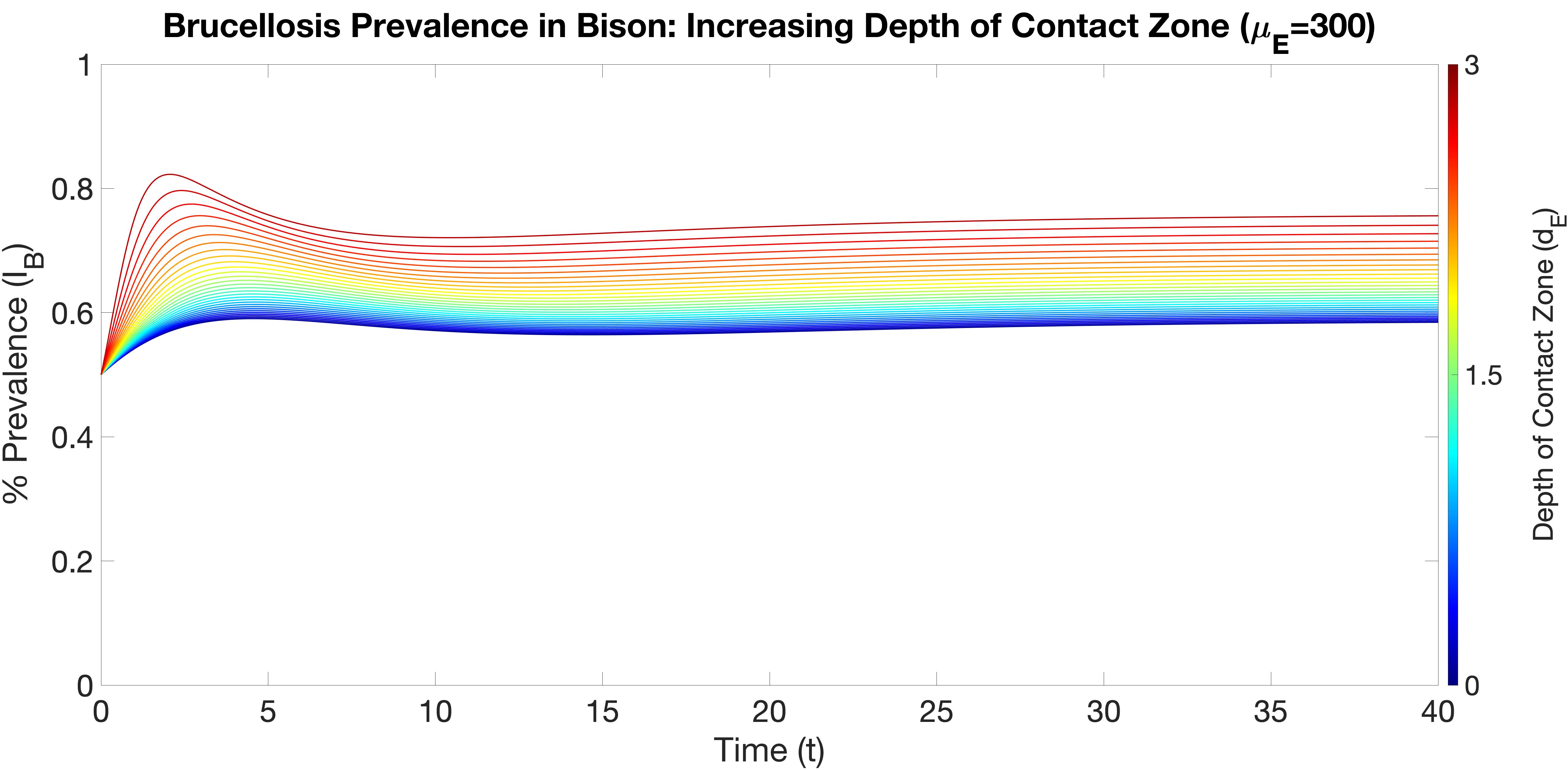}\label{THREE_CHANGE_D1_MU1_300_Bison}  }}

\caption
[(c) Bison Prevalence Curves as Change Depth of Contact Zone Between Cattle and Elk with High Shape Index]
{
(c) A time series plot showing the level of brucellosis prevalence in the bison population on a landscape where the contact zone between cattle and elk has the shape index of $(\mu_E=300)$ as the depth of the contact zone ($d_E$) increases from a minimum value of ($d_E=0$) (the darkest blue line) to an assigned value of ($d_E=3$) (the darkest red line).
}
\label{THREE_CHANGE_D1_MU1_300}
\end{figure}






\newpage

\begin{figure}[!htbp]
\centering
\subfloat[]{{\includegraphics[scale=.085]{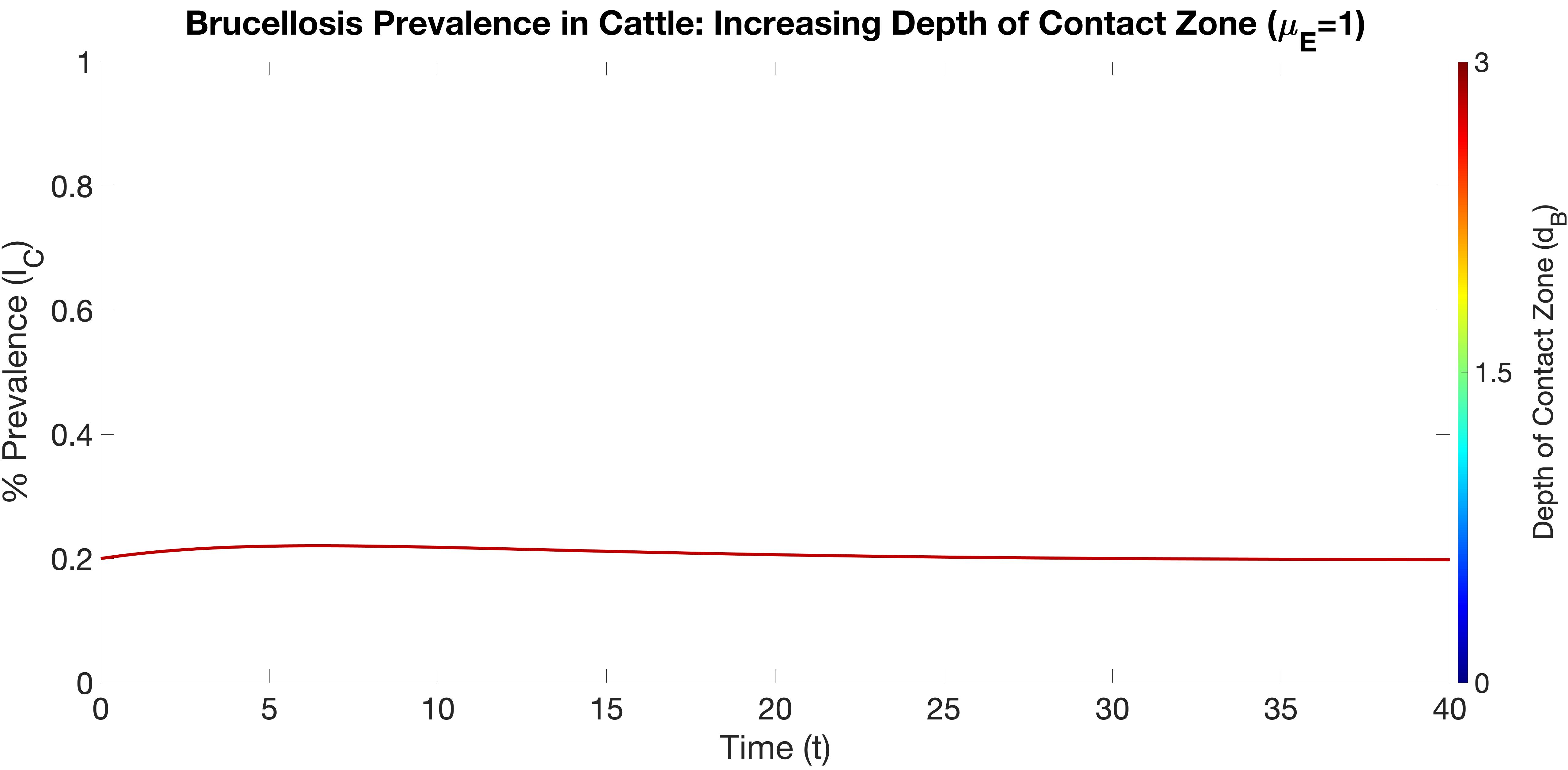}\label{THREE_CHANGE_D2_MU1_1_Cattle}   }}

%

\caption
[(a) Cattle Prevalence Curves as Change Depth of Contact Zone Between Elk and Bison with Low Shape Index]{
(a) A time series plot showing the level of brucellosis prevalence in the cattle population on a landscape where the contact zone between cattle and elk has the shape index of $(\mu_E=1)$ as the depth of the contact zone between elk and bison ($d_B$) increases from a minimum value of ($d_B=0$) to an assigned value of ($d_B=3$) (the darkest red line which overlays the other curves). 
}
\label{THREE_CHANGE_D2_MU1_1}
\end{figure}

\newpage

\begin{figure}[!htbp]
\ContinuedFloat
\centering

\subfloat[]{{\includegraphics[scale=.085]{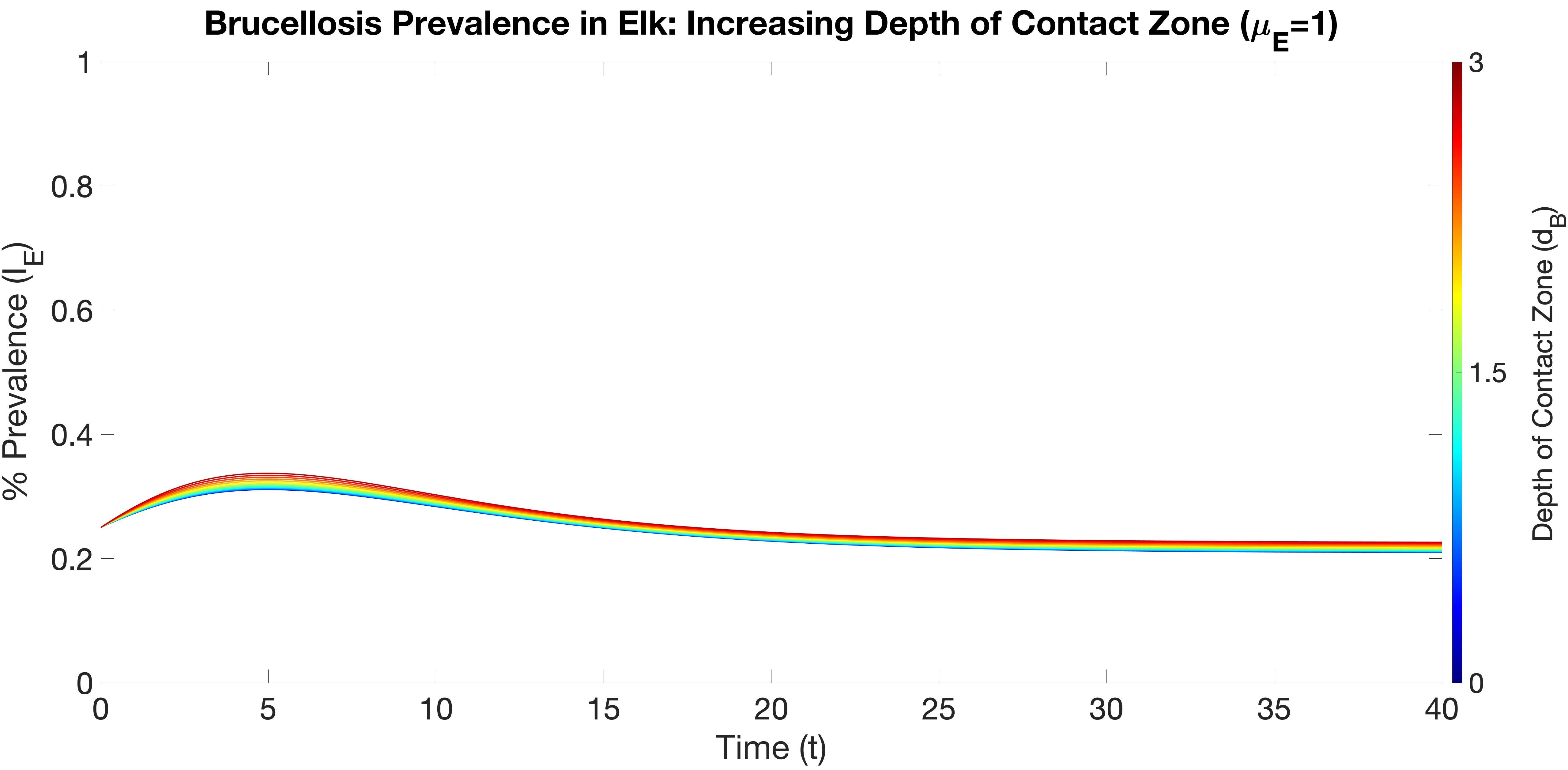}\label{THREE_CHANGE_D2_MU1_1_Elk}  }}


\caption
[(b) Elk Prevalence Curves as Change Depth of Contact Zone Between Elk and Bison with Low Shape Index]{\
(b) A time series plot showing the level of brucellosis prevalence in the elk population on a landscape where the contact zone between cattle and elk has the shape index of $(\mu_E=1)$ as the depth of the contact zone between elk and bison ($d_B$) increases from a minimum value of ($d_B=0$) (the darkest blue line) to an assigned value of ($d_B=3$) (the darkest red line). 
}
\label{THREE_CHANGE_D2_MU1_1}
\end{figure}

\newpage

\begin{figure}[!htbp]
\ContinuedFloat
\centering
%

\subfloat[]{{\includegraphics[scale=.085]{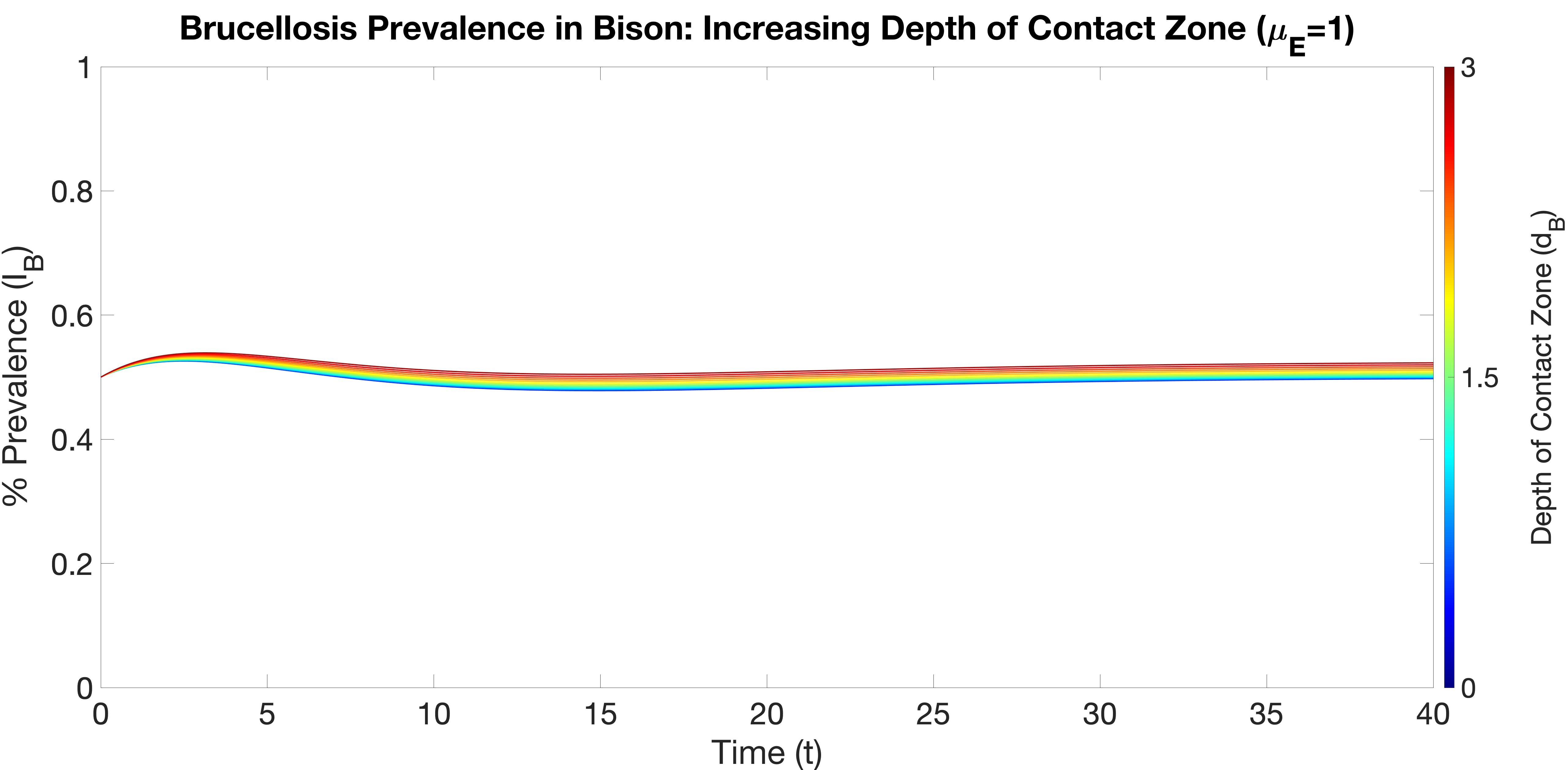}\label{THREE_CHANGE_D2_MU1_1_Bison}  }}

\caption
[(c) Bison Prevalence Curves as Change Depth of Contact Zone Between Elk and Bison with Low Shape Index]{
(c) A time series plot showing the level of brucellosis prevalence in the bison population on a landscape where the contact zone between cattle and elk has the shape index of $(\mu_E=1)$ as the depth of the contact zone between elk and bison ($d_B$) increases from a minimum value of ($d_B=0$) (the darkest blue line) to an assigned value of ($d_B=3$) (the darkest red line).
}
\label{THREE_CHANGE_D2_MU1_1}
\end{figure}


\newpage

\begin{figure}[!htbp]
\centering
\subfloat[]{{\includegraphics[scale=.085]{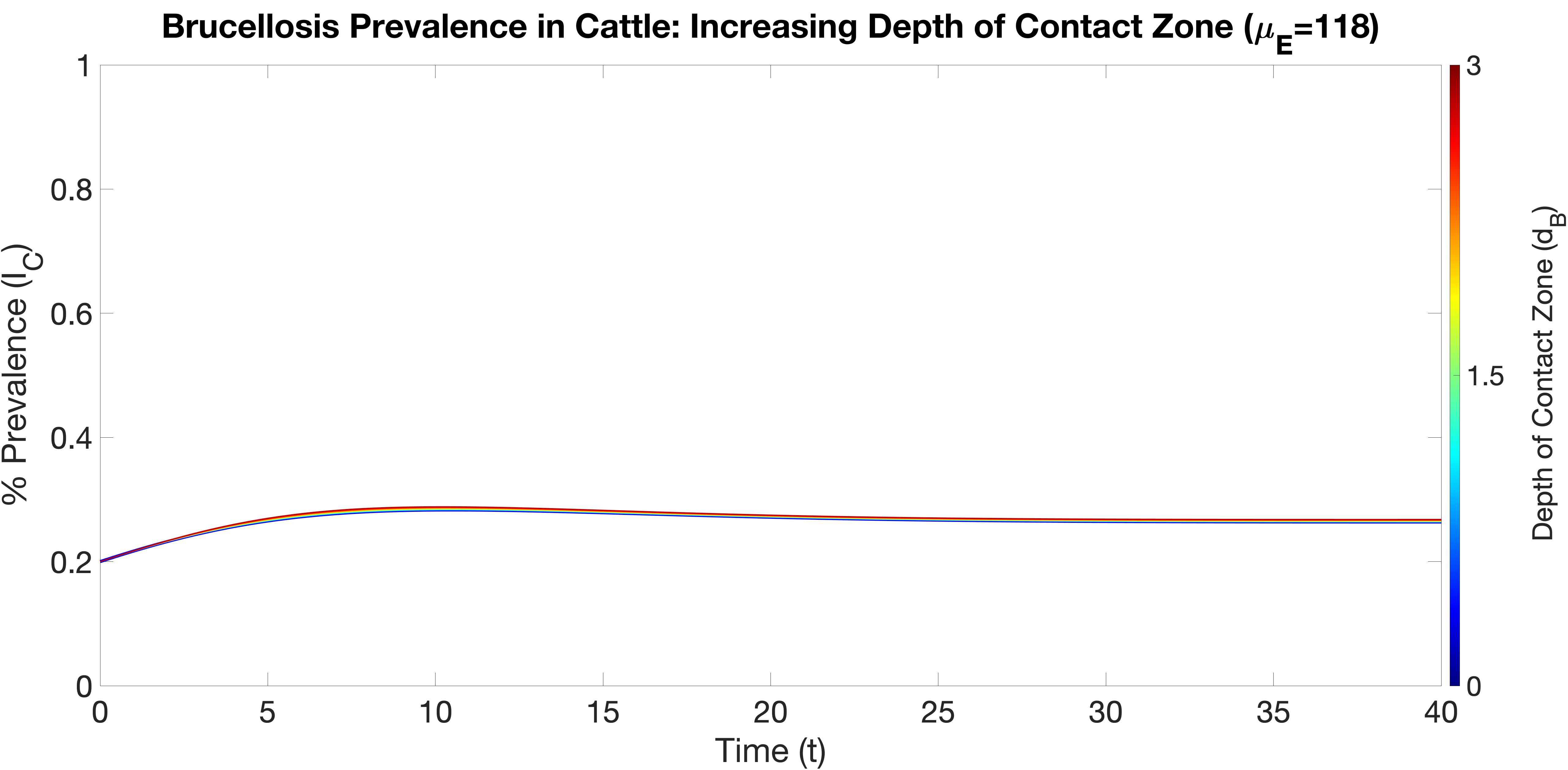}\label{THREE_CHANGE_D2_MU1_118_Cattle}   }}

%

\caption
[(a) Cattle Prevalence Curves as Change Depth of Contact Zone Between Elk and Bison with Medium Shape Index]{
(a) A time series plot showing the level of brucellosis prevalence in the cattle population on a landscape where the contact zone between cattle and elk has the shape index of $(\mu_E=118)$ as the depth of the contact zone between elk and bison ($d_B$) increases from a minimum value of ($d_B=0$) (the darkest blue line) to an assigned value of ($d_B=3$) (the darkest red line). 
}
\label{THREE_CHANGE_D2_MU1_118}
\end{figure}

\newpage

\begin{figure}[!htbp]
\ContinuedFloat
\centering

\subfloat[]{{\includegraphics[scale=.085]{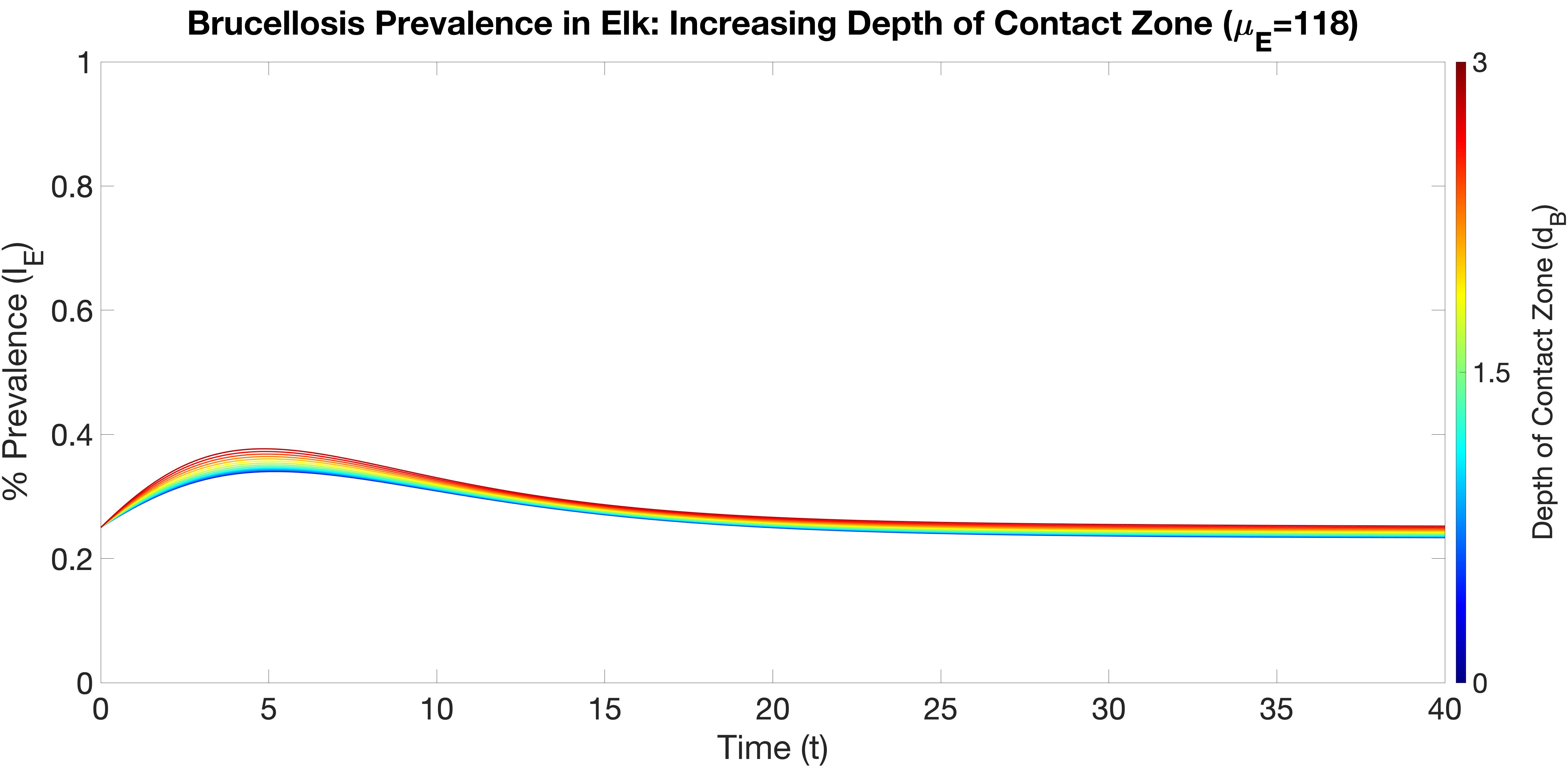}\label{THREE_CHANGE_D2_MU1_118_Elk}  }}


\caption
[(b) Elk Prevalence Curves as Change Depth of Contact Zone Between Elk and Bison with Medium Shape Index]{
(b) A time series plot showing the level of brucellosis prevalence in the elk population on a landscape where the contact zone between cattle and elk has the shape index of $(\mu_E=118)$ as the depth of the contact zone between elk and bison ($d_B$) increases from a minimum value of ($d_B=0$) (the darkest blue line) to an assigned value of ($d_B=3$) (the darkest red line). 
}
\label{THREE_CHANGE_D2_MU1_118}
\end{figure}

\newpage

\begin{figure}[!htbp]
\ContinuedFloat
\centering


\subfloat[]{{\includegraphics[scale=.085]{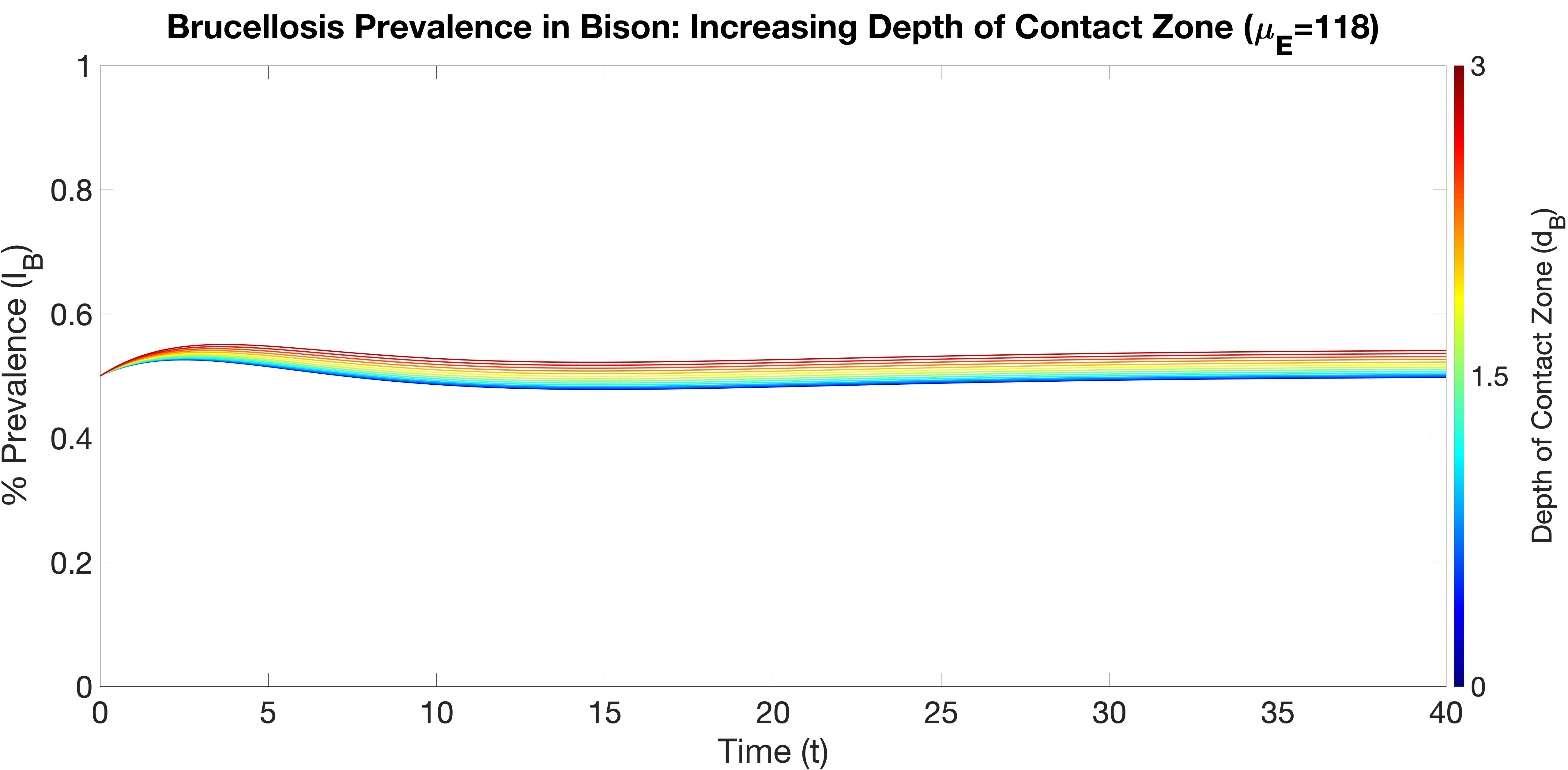}\label{THREE_CHANGE_D2_MU1_118_Bison}  }}

\caption
[(c) Bison Prevalence Curves as Change Depth of Contact Zone Between Elk and Bison with Medium Shape Index]{
(c) A time series plot showing the level of brucellosis prevalence in the bison population on a landscape where the contact zone between cattle and elk has the shape index of $(\mu_E=118)$ as the depth of the contact zone between elk and bison ($d_B$) increases from a minimum value of ($d_B=0$) (the darkest blue line) to an assigned value of ($d_B=3$) (the darkest red line).
}
\label{THREE_CHANGE_D2_MU1_118}
\end{figure}

\newpage

\begin{figure}[!http]
\centering
\subfloat[]{{\includegraphics[scale=.085]{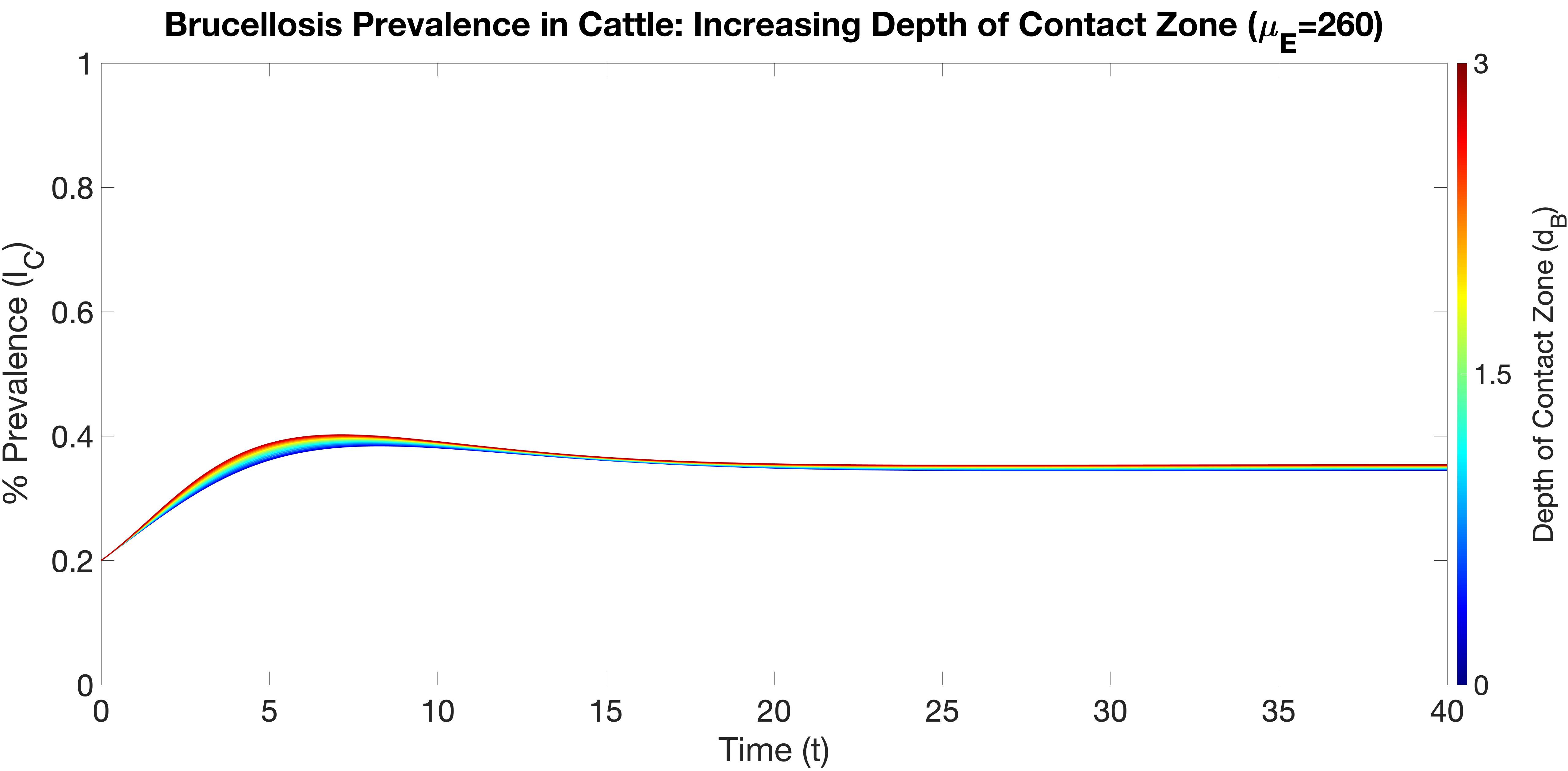}\label{THREE_CHANGE_D2_MU1_300_Cattle}   }}

%

\caption
[(a) Cattle Prevalence Curves as Change Depth of Contact Zone Between Elk and Bison with High Shape Index]
{
(a) A time series plot showing the level of brucellosis prevalence in the cattle population on a landscape where the contact zone between cattle and elk has the shape index of $(\mu_E=260)$ as the depth of the contact zone between elk and bison ($d_B$) increases from a minimum value of ($d_B=0$) (the darkest blue line) to an assigned value of ($d_B=3$) (the darkest red line). 
}
\end{figure}

\newpage

\begin{figure}[!http]
\ContinuedFloat
\centering

\subfloat[]{{\includegraphics[scale=.085]{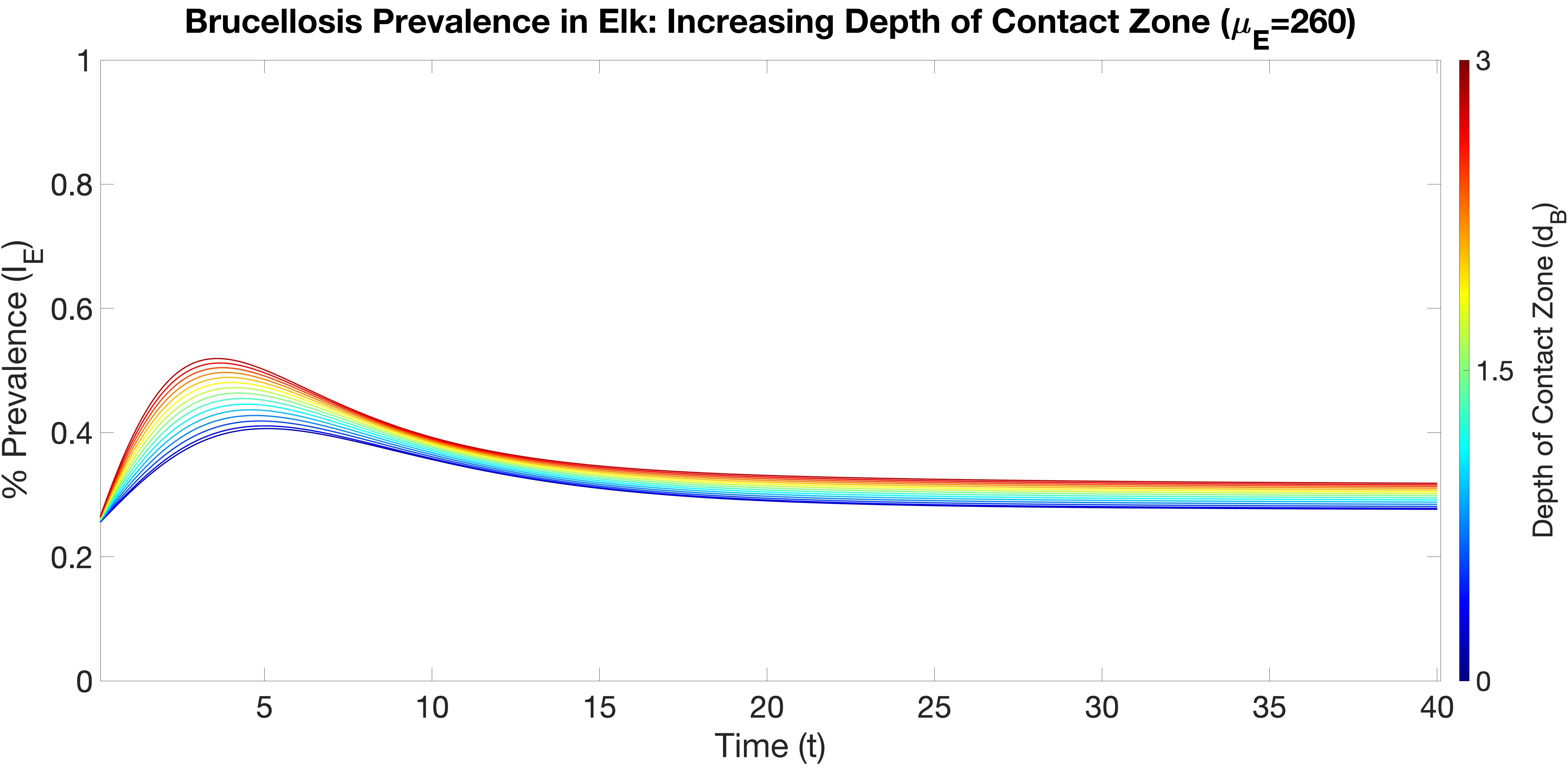}\label{THREE_CHANGE_D2_MU1_300_Elk}  }}


\caption
[(b) Elk Prevalence Curves as Change Depth of Contact Zone Between Elk and Bison with High Shape Index]
{
(b) A time series plot showing the level of brucellosis prevalence in the elk population on a landscape where the contact zone between cattle and elk has the shape index of $(\mu_E=260)$ as the depth of the contact zone between elk and bison ($d_B$) increases from a minimum value of ($d_B=0$) (the darkest blue line) to an assigned value of ($d_B=3$) (the darkest red line). 
}
\end{figure}

\newpage

\begin{figure}[!htb]
\ContinuedFloat
\centering

\subfloat[]{{\includegraphics[scale=.085]{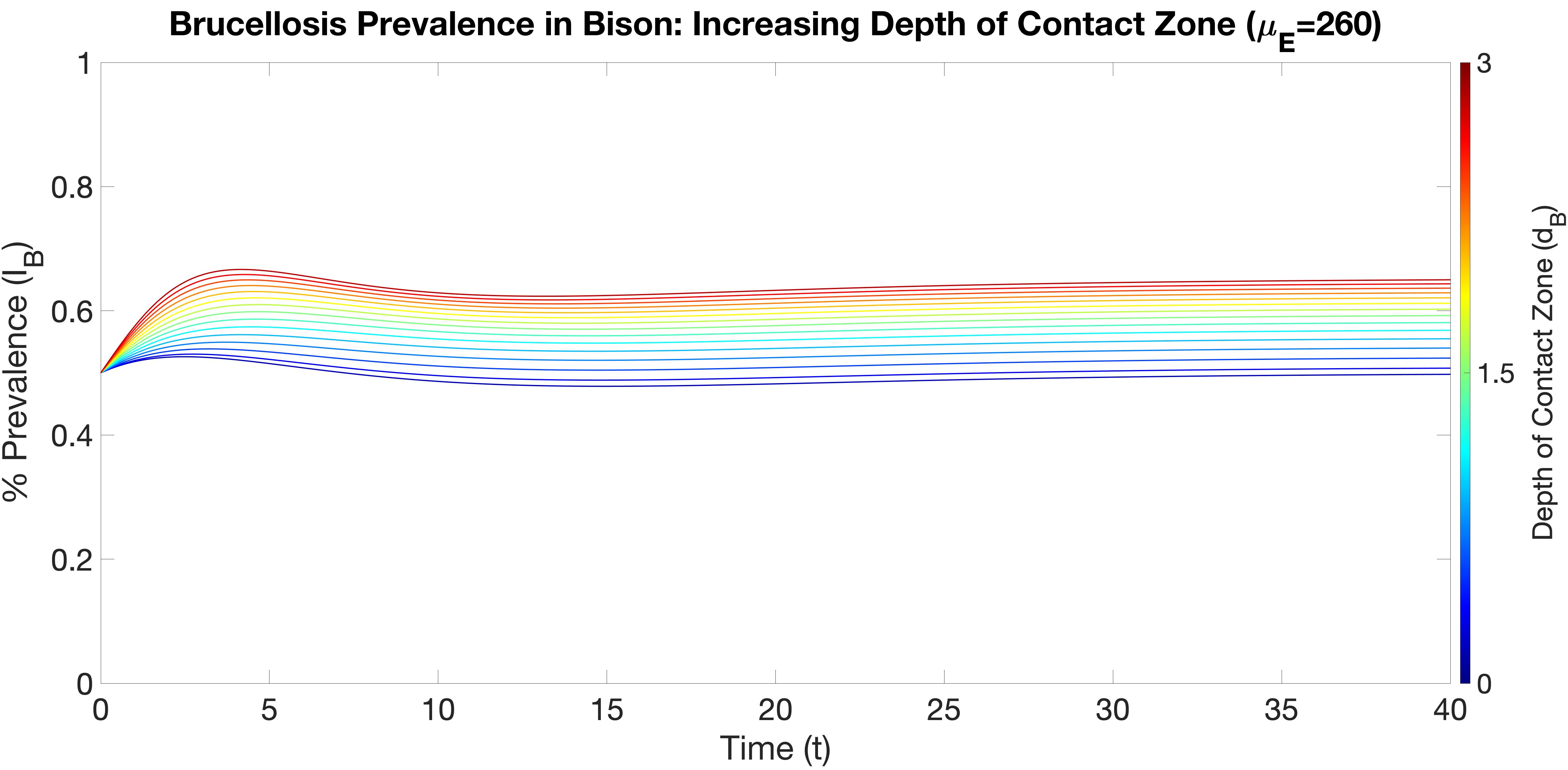}\label{THREE_CHANGE_D2_MU1_300_Bison}  }}

\caption
[(c) Bison Prevalence Curves as Change Depth of Contact Zone Between Elk and Bison with High Shape Index]{
(c) A time series plot showing the level of brucellosis prevalence in the bison population on a landscape where the contact zone between cattle and elk has the shape index of $(\mu_E=260)$ as the depth of the contact zone between elk and bison ($d_B$) increases from a minimum value of ($d_B=0$) (the darkest blue line) to an assigned value of ($d_B=3$) (the darkest red line).
}
\label{THREE_CHANGE_D2_MU1_300}
\end{figure}

%
%

\newpage

Figure (\ref{THREE_CHANGE_MU1}) shows that as the shape index of the habitat shared by elk and cattle changes (the shape index is increased), so that there is more edge (contact zone), there are increases in the initial rate of epidemic spread, peak prevalence, and endemic prevalence, not only in the elk and cattle populations, but also in the bison population that is not situated on the segment of the ecosystem where land-use change occurs. 

When the landscape between elk and cattle is altered -- by changing the depth of the contact zone $(d_E)$, there are discernible changes to the brucellosis prevalence in all species for different shape indices. For Figure (\ref{THREE_CHANGE_D1_MU1_1}), Figure (\ref{THREE_CHANGE_D1_MU1_118}), and Figure (\ref{THREE_CHANGE_D1_MU1_300}), the depth of the contact zone between cattle and elk increases from 0 (indicated by the darkest blue line) to 3 meters (indicated by the darkest red line). In Figure (\ref{THREE_CHANGE_D1_MU1_1}), the shape index of the contact zone is $(\mu_E=1)$, and as $(d_E)$ increases, there is not a significant change of the initial rate of epidemic spread, peak prevalence, or endemic prevalence in the cattle, elk, and bison populations. In Figure (\ref{THREE_CHANGE_D1_MU1_118}) and Figure (\ref{THREE_CHANGE_D1_MU1_300}), the shape index of the contact zone is ($\mu_E=118$) and ($\mu_E=300$), respectively. As $(d_E)$ increases, there is an increase in the initial rate of epidemic spread, peak prevalence, and endemic prevalence in all three populations. In the bison population, the prevalence change is significantly smaller than the other species in Figure (\ref{THREE_CHANGE_D1_MU1_118}), but there is an observable increase in Figure (\ref{THREE_CHANGE_D1_MU1_300}) compared to Figure (\ref{THREE_CHANGE_D1_MU1_1}) and Figure (\ref{THREE_CHANGE_D1_MU1_118}).

Another question is then: What happens to the brucellosis prevalence in cattle as the landscape between elk and bison is altered? The changes of bison and elk interactions in their overlap region can impact disease prevalence in the non-overlapping cattle population. In Figure (\ref{THREE_CHANGE_D2_MU1_1}), the habitat overlap between cattle and elk has the shape index of $(\mu_E=1)$. As the depth of the contact zone between elk and bison $(d_B = 0)$ increases from a minimum value of $(d_B = 0)$ (indicated by the darkest blue curve) to $(d_B = 3)$ meters (indicated by the darkest red curve), there is no change in the initial rate of epidemic spread, peak prevalence, or the endemic prevalence in the cattle population. There is, however, an increase in all of these epidemic characteristics in 

In Figure (\ref{THREE_CHANGE_D2_MU1_118}) and Figure (\ref{THREE_CHANGE_D2_MU1_300}), the shape index of the habitat overlap between cattle and elk are $(\mu_E=118)$ and $(\mu_E=260)$, respectively. For both figures, the initial rate of epidemic spread,  peak prevalence, and the endemic prevalence increase in all three species as the depth of the contact zone between elk and bison $(d_B = 0)$ increases from a minimum value of $(d_B = 0)$ (indicated by the darkest blue curve) to $(d_B = 3)$ meters (indicated by the darkest red curve). There is an even greater increase in the epidemic characteristics for $(\mu_E=260)$.

\section{Discussion}

In response to the \citet{national2017revisiting}, which recommends to characterize the risk brucellosis transmission from elk, and assess drivers of land-use change and their effects on disease spread, a model was constructed that had not been applied to this system before. Specifically, as an extension of the work by \citet{dobson1996population}, this approach is the first to incorporate bison, elk, and cattle into a model for brucellosis transmission in the GYE. Moreover, using the findings of \citet{rayl2019modeling}, which provided an understanding of the spatial processes involved in the epidemiology of the disease, landscape ecology metrics were incorporated to estimate how habitat overlap and land-use change facilitate disease transmission. Epidemiological parameters were taken from literature, and the simulations approximated disease prevalence in all three populations as the contact zone between two species was adjusted. The estimated prevalence levels, assuming the assumptions of the system do not change dramatically, can be used as predictions. Moreover, the prevalence levels of the elk population correspond with the results, of \citet{cotterill2018winter}, but expands the literature by also providing prevalence estimates for elk and cattle and associating those estimates with land-use change. The results have implications for brucellosis management in the GYE and also contributes to the literature of ecology and epidemiology.

It was found that incorporating bison did not change them main results from the previous chapter, but showed how landscape changes in particular regions of the GYE might be expected to impact species elsewhere in the region. The numerical experiments indicate that increasing the depth of the contact zone between elk and bison ($d_b$), the shape of the contact zone between cattle and elk ($\mu_E$), and the depth of the contact zone between cattle and elk ($d_E$) causes an increase in the initial rate of epidemic spread, peak prevalence, and endemic prevalence, in both the species that are situated where the landscape change occurs and in those not situated where the change occurs. It was found that increased contact between elk and cattle translates to more brucellosis prevalence in the bison population. Likewise, it was found that increased contact between elk and bison translates to more brucellosis prevalence in the cattle population. 

Increasing the shape index of the cattle and elk habitat overlap ($\mu_E$) caused an increase of brucellosis prevalence in cattle, elk, and bison. As the shape index ($\mu_E$) increased, there was an increase in the amount of edge between habitat types. This is an increase in the amount of habitat overlap between cattle and elk, resulting in more species interactions, and, thus, more disease transmission between these species. As elk are the intermediary for disease transmission between cattle and bison, an increase in cattle and elk habitat overlap shape index ($\mu_E$) will also result in an increase in the brucellosis prevalence in bison.
\newpage

Perturbations to ($\mu_E$) can be attributed to altered migration patterns of elk, which could result from factors affecting the availability of resources for elk, causing changes in elk foraging. These factors could be fencing, the instigation of new or more cattle production to the GYE, and other forms of land-use change in the region. Fencing may restrict elk migration and limit their access to resources, and increased cattle herd density may deplete resources, both of which cause the elk to forage in other locations. Moreover, an increase in the amount of winter feedgrounds for elk that are distributed in a spatially sparse manner could cause a shift in their congregations; consolidating where elk inhabit would adjust the shape of their foraging range. Additionally, as more sections of the GYE are established for cattle production, the landscape could become more habitable for elk and cause elk to forage in new places. This would impact ($\mu_E$) more if the sectioning was especially sparsely distributed.

The depth of the contact zone between cattle and elk is ($d_E$). With an increase of ($d_E$), the amount of edge between habitat types increases, thus an increase of habitat overlap between the respective species. This results in more interactions and brucellosis prevalence in cattle and elk. An increase in cattle and elk habitat overlap depth ($d_E$) will also result in an increase in the brucellosis prevalence in bison, since elk are an intermediary for disease transmission between cattle and bison.

To decrease brucellosis transmission in bison, elk, and cattle in the GYE, species interactions must decrease; an increase of species interactions may be due to an increase of habitat overlap. 
Moreover, an expansion of cattle ranching operations could increase the habitat overlap between cattle and elk. Generally, cattle in this region are located on lowland grasslands, and the foraging preference of elk could increase the amount of habitat overlap, as both show a preference to the same type of habitat. Nevertheless, land-use changes have increased the proximity of these species \newpage \noindent together and as such there has been a sustained amount of spillover cases to cattle \citep{hansen2009species}.

The depth of the contact zone between elk and bison is ($d_B$). With an increase of ($d_B$), the amount of edge between habitat types increases, thus an increase of habitat overlap between the respective species. This results in more interactions and brucellosis prevalence in elk and bison. An increase in elk and bison habitat overlap depth ($d_B$) will also result in an increase in the brucellosis prevalence in cattle, since elk are an intermediary for disease transmission between cattle and bison.


This may be due to changes in migration patterns of elk into the national parks. Changes to ($d_b$) may reflect changes in foraging preference and migration patterns of elk into the habitat of bison, ($a_b$), which represents the national parks. Since a large shift in land-use change is currently not permitted on the national parks, a factor that will not impact ($d_b$) is in land-use change directly applied to ($a_b$). The numerical experiments shown in Figure  (\ref{THREE_CHANGE_D2_MU1_1}), Figure (\ref{THREE_CHANGE_D2_MU1_118}), and Figure (\ref{THREE_CHANGE_D2_MU1_300}) emphasize that since elk migrate and mix with bison in the national parks, they act as an intermediary for disease transmission from bison to cattle, and the influx of land-use change thought the GYE has facilitated an increase of brucellosis prevalence in all three species. These results have implications for brucellosis management in the GYE.

\subsection{Management Strategies}

Brucellosis is detected within all regions that are inhabited by cattle, elk, and bison in the GYE, and, as a consequence, coextends to their respective habitats \citep{national2017revisiting,proffitt2011elk}. As elk and bison are both reservoirs for brucellosis, and elk act as an intermediary for disease transmission between cattle and bison, management efforts need to be devoted towards all three species (thus in all habitats) in order to eliminate spillover cases to cattle in the region \citep{rhyan2009pathogenesis,white2011management}. Although the current strategies of cattle vaccination and the segregation of cattle and bison have been effective in reducing spillover cases to cattle, no single management strategy is going to eliminate brucellosis from cattle in the area \citep{national2017revisiting}. The results of the model suggest that land-use change has contributed to an increase in disease prevalence, and that habitat management could be used to control the disease. The model shows how habitat management that changes parameters ($d_B$), ($\mu_E$), and ($d_E$) may be used to control disease. 

 In the model, the parameter ($d_b$) is the depth of the contact zone between elk and bison, and as ($d_b$) increases, the species' interactions increase. Realistically, elk are allowed to forage onto all of the habitat of bison, as ($d_b$) is an abstraction of elk movements into the national parks. In terms of disease spread, the ideal result of management solutions (related to ($d_b$)) is to minimize the contact between elk and bison. A possible solution may be, though extreme, to build a wildlife fence or buffer zone around YNP and GTNP. Bison would continue to be confined inside the parks, while elk would restricted from entering and migrating.

In the model, the parameter ($d_b$) is the depth of the contact zone between elk and bison. As ($d_b$) increases, interactions between species also increases. In principle, elk have access to all of the habitat of bison. In practice, ($d_b$) is limited by elk migration patterns. In terms of disease spread, ($d_b$) is a proxy for contact between elk and bison. The construction of a veterinary cordon fence between elk and bison would minimize ($d_b$). A fence around YNP and GTNP, for example, would confine bison inside the parks, and exclude elk from entering. Creating a buffer zone or wildlife fence was in fact proposed by \citet{dobson1996population}, but without considering how disruption to the migration patterns of elk could affect the ecosystem.
It is beyond the scope of this paper to recommend particular management strategies, but it is worth identifying strategies that work on one or other of these parameters. ($d_b$) might also be changed by measures to lessen contact between elk and bison through their spatial-temporal when both occupy the national parks. Even though elk are allowed to free range throughout the region, efforts to deter them from foraging in the same location at the same time as bison could reduce disease transmission, and, as the model suggests, would ultimately result in a decreased amount of spillover-cases to cattle.

%

Similar to ($d_b$), the shape of the contact zone between elk and cattle ($\mu_E$) can be a result of management actions taken. Changes to ($\mu_E$) can be attributed to altered migration patterns of elk. A possible land management strategy that could alter migration routes of elk may be the temporary removal of the supplemental winter feedgrounds. The removal of the winter feedground supports the findings of \citet{cotterill2018winter,brennan2017shifting}. They found that supplementally feeding elk at the feedgrounds during the winter increases the brucellosis prevalence in their population and contributes to spillover cases in cattle. Moreover, although the feedgrounds were established to deter elk from interacting with cattle, their placement seems to be directing elk to migrate in particular directions and to congregate in particular regions; this results in foraging range that has a noncircular shape and increases the amount of edge between the habitat of elk and cattle. The suspension of the feed grounds, at least until brucellosis is eradicated, could force elk to develop a more circular foraging range. Ecologically, a reduction of edge between habitat types results in less disease spread. In the confines of the model, reducing muE, reduces the amount of habitat edge that the elk share with cattle. Thus, within the model, it is beneficial that elk migrate in a more scattered, homogenous manner.

In theory, reducing of the edge to area ratio of elk habitat would reduce the likelihood of cross-species infection. However, the simulations did not account for an increase in elk habitat area that may result from an expansion of foraging, as the numerical experiments assumed the area was held constant while the perimeter increased. Hence, as elk migrate during the winter season to lowland grasslands where they can interact with cattle, if the feedgounds are removed and they are no longer congregating in particular regions, the elk would be expected to have more interaction with cattle. Thus, in order to diminish the likelihood of cross-species transmission, cattle and elk may also need to be separated at the lowland grasslands in the winter season.
 
In conjunction with reducing ($\mu_E$), the model indicates that if efforts to reduce the level of brucellosis prevalence throughout the ecosystem may be achieved by mitigating ($d_E$) through land management techniques as well. By restricting cattle movements, through methods such as placing them into feedlots, during the season that brucellosis spillovers are most likely, there can be a reduction in disease transmission. The control efforts exhibited to all species may serve in eradicating the disease, not only from cattle but also in the wildlife. Though the results have implications for brucellosis management in the GYE, this work also contributes to the literature of ecology and epidemiology.

\section{Conclusion}

Broadly, this study addressed the direct and indirect consequences of landscape management for the control of epizootic disease spread between species. The indirect consequences include effects on species other than those occupying the segment of land that is modified. The focus on brucellosis transmission in the GYE was motivated by the \citet{national2017revisiting} report . The objective of the study was to model brucellosis transmission dynamics between bison, elk, and cattle. To approach this, a system of differential equations was developed and landscape ecology metrics were incorporated to determine how habitat overlap and land-use change facilitate disease spread, especially for a species not situated where the land-use change occurs.

\citet{national2017revisiting} report encouraged work to characterize the risk of brucellosis transmission to cattle and bison from elk and to assess the effects that land-use change has on disease spread. This study contributes a novel model of \textit{B. abortus} transmission in the GYE, to estimate disease prevalence in cattle, elk, and bison as the contact zone between either cattle and elk or elk and bison was adjusted. It was determined that landscape alteration in one region could impact the disease prevalence of a species not situated there. As the depth of contact zone and shape of the habitat overlap for one region were increased, there was an increase in prevalence not only in the species where the landscape alteration occurred, but also in the species not situated there, assuming some species acts as an intermediary for disease transmission.


Brucellosis in the GYE continues to be a persistent issue for cattle ranchers and wildlife management agencies \citep{ragan2002animal}. Land-use change has facilitated higher disease prevalence throughout the GYE and continues to contribute to spillover-cases in cattle \citep{national2017revisiting}. The approach developed in the study may inform cattle ranchers interested in managing their private lands to reduce species interactions and thus spillover-cases to their livestock.

In regards to the bison, the approach of this study may also inform the IBMP as to how land-use change might be driving higher brucellosis prevalence in the Yellowstone bison. Since they and other conservationist groups care about the establishment of habitat for bison outside of the national parks, their efforts may need to encompass more than just the bison population. With the findings that land-use change, in conjunction with elk migration patterns, have contributed to the spread of brucellosis, the USFS, in particular, may take insight from this study to make land management decisions that can reduce disease spill-overs to cattle and transmission to bison, since elk heavily migrate throughout the national forests between the national parks and private land.  The results have implications in the context of brucellosis in the GYE and in the larger body of landscape epidemiology. 

The results also have implications for the larger body of landscape epidemiology, where there are studies that seem to contradict as well as studies that support the conclusions of this research. Conflicting literature concludes that more anthropogenic habitat fragmentation (i.e. more land-use change) either isolates populations to the extent that there is a decrease in species interactions, and thus enzootic contagion, or reduces the biodiversity in an ecosystem so that particular diseases' transmission becomes amplified, especially at intermediate levels of land conversion \citep{faust2018pathogen,tracey2014agent,dion2011landscape}. Both of these conclusions contradict the finding that more land-use change leads to increased cross-species disease spread \citep{cantrell2001brucellosis,rulli2017nexus}. The conflicting results come from underlying assumptions made in each case about how the edge impacts species interactions. The model developed in this chapter assumes that there is more homogenous mixing as more edge develops, and that an absence of edge results in the same conclusion. The optimal result is to have some edge but for it to be organized in a manner that minimizes the contact between habitat types. The assumptions are derived from the work of \citep{dobson1996population,murcia1995edge,patton1975diversity} and support the work of \citep{game1980best,johnson2011edge}. 

The novelty of this model is that it examines disease transmission mediated by a species between two other unconnected species. Moreover, it accounts for how habitat shapes and depth of contact zones contribute to disease spread. The implication of these ecological factors in the field of landscape epidemiology is that to improve the accuracy of mathematical model utilized, it is imperative to consider more disease reservoirs and their associated environmental conditions. This will facilitate comprehension of disease propagation and its second order effects to species in the landscape as land-use changes and policies are implemented. Consequently, these models should not be applied to a global scale, but applied to specific ecosystems.

In terms of the application, future work could incorporate vertical transmission, compare these results to data, and include more rigorous parameter estimation techniques. Vaccination, seasonality, and indirect infection variables could also be considered in order to improve the accuracy of the results. Nevertheless, this study's contributions reveal more of the true nature of brucellosis transmission in the GYE.

\bibliographystyle{spbasic}      

\bibliography{citations.bib}

\end{document}